# Topological Pharmacokinetics: A Model-Free Framework for Quantifying the Shape of Concentration–Time Curves

Authors: Hong-Can Ren[1], Ya-Xin Gu[1]

[1] *Changchun GeneScience Pharmaceutical Co., Ltd. (GeneScience Pharmaceuticals Co., Ltd.), Changchun, China*

Corresponding authors. Hong-Can Ren, E-mail: renhongcan@163.com

## Abstract

Pharmacokinetic analysis compresses concentration–time curves into scalar summaries (non-compartmental analysis, NCA) or fits pre-specified compartmental models—the former being model-free but shape-blind, the latter shape-sensitive but structure-dependent. We introduce Topological Pharmacokinetics (TPK), a framework that reads the shape of concentration–time curves directly from data without prior commitment to any model. TPK delay-embeds each curve into a state-space trajectory and applies persistent homology to extract a nine-component pharmacokinetic topological invariant (PTI) vector: multi-scale shape descriptors—peak hierarchy, secondary-peak isolation, loop structure—derived from persistence diagrams with established perturbation-stability properties. We evaluate the framework on four layers of escalating realism. Four monophasic linear systems generate no spurious cyclic signals, whereas double-site absorption and enterohepatic recirculation produce distinct loop signatures; 12 nonlinear scenarios, each paired with its linear counterpart, resolve as curvature-without-cycles, with linear shape rigidity contrasted against dose-dependent deformation; a weak endogenous rhythm survives superposition with a large exogenous drug signal (shape-deviation statistic=1.410, inter-peak interval recovered at exactly 24 h); and a systematic operating-characteristics analysis maps the envelope—loop detection stable to 10% of $C_{max}$ additive noise and to sampling thinned to 7 points for peak-structure cycles, bracketing the clinical sparse-sampling regime. As clinical proof of concept, we apply the identical pipeline to 15 real mycophenolic acid profiles digitized from a published kidney-transplant study. The secondary-persistence ratio is associated with ex vivo gut-microbiome reactivation rates at Spearman $\rho = 0.815$ ($p = 0.0002$, $q = 0.002$; stable across parameter grids, exclusions, and worst-case effective-n = 10), where the tested exposure metrics show weak, nonsignificant associations ($\rho \leq 0.38$); exploratory partial-correlation analysis suggests information beyond the tested conventional recirculation metrics. Persistence encodes peak isolation—a multi-scale property that neither rebound height nor operator-dependent AUC ratios capture. All analyses are fully automated and reproducible; code and data are public. TPK is proposed as a complementary, model-free, shape-sensitive layer alongside NCA and model-based pharmacokinetics—the two-dimensional foundation of a program extending to topological pharmacodynamics.

## Study Highlights

### What is the current knowledge on the topic?

Pharmacokinetic analysis reduces concentration–time curves to scalar summaries (AUC, $C_{max}$) or fits pre-specified compartmental models; shape information that encodes processes such as enterohepatic recirculation, nonlinear elimination, and endogenous rhythms is systematically discarded, and its clinical quantification (e.g., EHR AUC ratios) is operator-dependent.

### What question did this study address?

Can the shape of a concentration–time curve be quantified directly from data—without pre-committing to a compartmental model—robustly enough for sparse clinical sampling, and does it carry information that existing methods do not extract?

**What does this study add to our knowledge?**

A model-free, shape-sensitive framework (delay embedding + persistent homology → PTI vector) with a four-layer evaluation: no spurious loops in four monophasic linear controls; nonlinearity resolved as curvature-without-cycles; endogenous rhythms detected through superposition-invariant shape statistics; and, on 15 real MPA profiles, secondary-peak topology associated with gut-microbiome reactivation ($\rho = 0.815$) where the tested exposure metrics show weak, nonsignificant associations ($\rho \leq 0.38$), with the operating envelope (noise, sampling density, embedding parameters) mapped systematically.

**How might this change drug discovery, development, and/or therapeutics?**

TPK converts the discarded shape of concentration–time curves into a directly computable covariate for exposure–response analysis and regimen design. Shape is the temporal structure of exposure: for narrow-therapeutic-index drugs it can carry efficacy and toxicity information that AUC and $C_{max}$ average away; automated per-patient shape phenotypes inform dosing frequency, food state, interaction management, and monitoring intensity; and a fully automated sentinel flags nonlinear elimination, rhythmic interference, and anomalous secondary structure in early development and in already-published data. In each use the endpoint is the one pharmacokinetics ultimately serves: efficacy and safety in the individual patient.

## 1. Introduction

Pharmacokinetic inference rests on two complementary frameworks, each with a structural blind spot. Non-compartmental analysis (NCA) summarizes exposure—AUC, $C_{max}$, half-life—robustly and without model assumptions, but discards the shape of the concentration–time curve by construction [1,2]. Compartmental and population pharmacokinetic models do capture shape, but only through an assumed parametric structure: a sum of exponentials sees the world as sums of exponentials, and any shape outside the pre-selected function family is relegated to the residual error term [3,4,5,6]. There is thus a vacant position in the methodological landscape: an analysis that is simultaneously model-free *and* shape-sensitive—general like NCA, shape-aware like compartmental modeling, but committed to neither.

The vacancy is consequential because shape is often precisely where the signal resides. Enterohepatic recirculation (EHR) produces a secondary peak 6–10 h after dosing; saturable elimination bows the terminal phase away from log-linearity; endogenous circadian rhythms superimpose periodic oscillations on the declining curve [7,8,9]. Each of these departures is a mechanistic fingerprint, and each is systematically averaged away by scalar summaries or absorbed into residuals by misspecified models. The canonical example is mycophenolic acid (MPA): its EHR secondary peak varies widely among patients, and a recent kidney-transplant study attributed this variability to gut-microbiome β-glucuronidase reactivation of biliary glucuronide [10]. Decisively, reactivation rates correlated with the degree of recirculation ($\rho = 0.76$) but not with AUC or trough concentrations ($\rho = 0.30$): the microbiome signal lives in trajectory *shape*, not exposure—largely invisible to NCA's coordinate system, and quantified in the source study by an operator-dependent, manually keyed AUC ratio.

Topological data analysis offers exactly the missing instrument. Persistent homology—the core engine of TDA—tracks geometric features (connected components, loops) as they appear, persist, and disappear across a multiscale filtration of the data, extracting shape descriptors without parametric assumptions [11,12,13]. Long-lived features correspond to genuine structure; short-lived ones are attributed to noise—a separation guaranteed by the stability theory of persistence [14,15]. TDA has characterized neuronal branching, gene-regulatory dynamics, single-cell architecture, and protein conformational landscapes [16,17,18,19]. Beyond these characterization studies, TDA has identified a survival-distinguishing breast-cancer subgroup [20], extracted shape-based structure from large genomic surveys [21], revealed dynamical organization of the working brain [22], and analyzed heart-rate variability [23]. Pharmacokinetics is not merely another application domain: it is a domain in which the geometry of the data is the direct imprint of the physiology that generated it, and in which reading shape without a model fills a blind spot that has persisted for half a century.

We introduce **Topological Pharmacokinetics (TPK)**. A concentration–time trajectory is treated as the observable projection of a dynamical system—absorption, distribution, metabolism, excretion—evolving on a low-dimensional manifold. Delay-coordinate embedding, justified generically by Takens' theorem [24,25,26], reconstructs a point cloud from the scalar curve alone; persistent homology then extracts its multi-scale shape invariants—connected components and loops—as a **pharmacokinetic topological invariant (PTI) vector**: a fixed-length, model-agnostic shape fingerprint, computed by a fixed pipeline with no model-selection step, and therefore a standardized, machine-readable feature set suitable for large-scale pharmacokinetic data integration.

This article develops TPK fully—theory, validation, and a clinical proof of concept—in a four-layer progression of escalating realism (Fig 1; Table 1). **Layer 1** establishes the negative control: four monophasic linear systems must generate no spurious cyclic signals, while double-site absorption and enterohepatic recirculation should retain their mechanism-distinct loop signatures. **Layer 2** asks whether nonlinearity is resolved as shape deformation: a full factorial of 24 scenarios—12 Michaelis–Menten systems, each paired with its linear counterpart. **Layer 3** asks whether a weak endogenous rhythm survives superposition with a large exogenous drug signal. **Layer 4** is the clinical test: 15 real MPA profiles—sparse, irregularly sampled, digitized from published data—analyzed by the identical pipeline, with the gut microbiome as an external anchor. The operating characteristics of the method (noise, sampling density, embedding parameters) are mapped systematically, so that every application can verify its data regime falls within the envelope. Throughout, TPK is positioned as a complementary layer—alongside, never replacing—NCA, population PK, and PBPK modeling. This article is the foundational, two-dimensional installment of a program whose subsequent steps extend the framework to higher-dimensional embeddings and to pharmacodynamics (Topological Pharmacology).

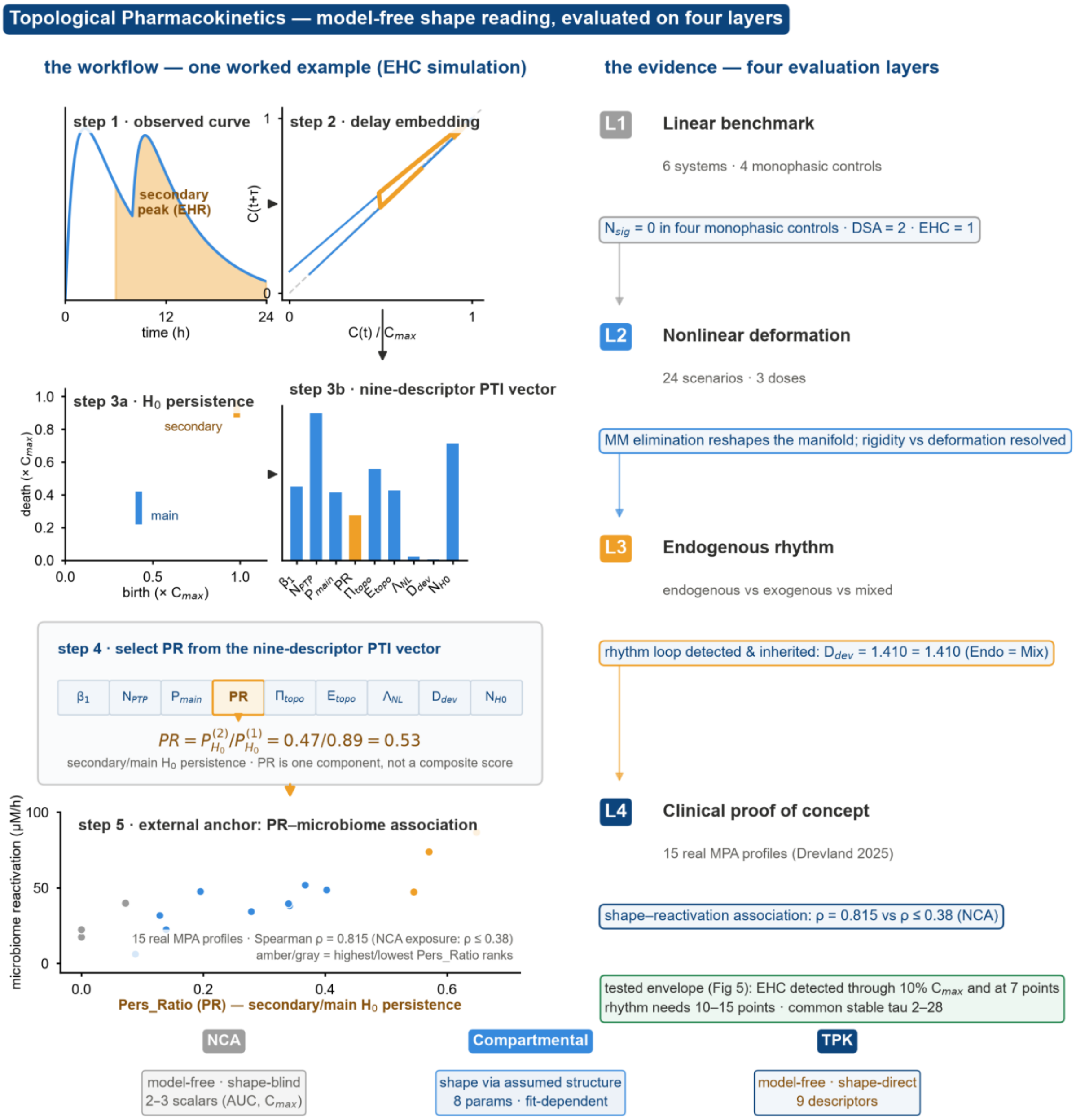

**Fig 1. Topological Pharmacokinetics at a glance: a model-free shape reader, evaluated on four layers.** Left: the workflow, carried by one worked example (a simulated enterohepatic-recirculation profile)—(step 1) the observed concentration–time curve, with the secondary peak shaded amber; (step 2) delay-coordinate embedding into state space, where recirculation closes a loop; (step 3a) H0 persistence, where bar length equals peak prominence; (step 3b) the nine-component PTI vector; (step 4) selection of Pers_Ratio (PR), one PTI component rather than a composite score, defined as secondary/main H0 persistence and illustrated by the worked-example calculation; (step 5) external anchoring on real data—PR computed for each of the 15 MPA profiles is associated with microbiome reactivation (Spearman $\rho = 0.815$; NCA exposure metrics $\rho \leq 0.38$; amber/gray = highest/lowest Pers_Ratio ranks). Right: the evidence ladder—one line per evaluation layer, each with its one-sentence result; the L4 line carries the value proposition ($\rho = 0.815$ vs $\rho \leq 0.38$). Bottom: the three-method positioning—NCA (model-free, shape-blind), compartmental fitting (shape through assumed structure), and TPK (model-free, shape-direct).

**Table 1.** The four-layer evaluation at a glance: one row per layer—system, ground truth, core result, and what it proves.

| Layer | System | Ground truth | Core result | Interpretation |
|---|---|---|---|---|
| L1 | six linear systems (incl. DSA, EHC) | no spurious cycles | N_sig = 0 in four monophasic systems; DSA 2, EHC 1 (mechanism-distinct signatures) | no spurious loops in monophasic controls; stratified double-peak descriptions |
| L2 | 24 scenarios (12 MM + 12 linear) | curvature ≠ cycles | N_sig = 0 in 24/24; linear shape rigidity vs | nonlinearity resolved without any model |
| | | | MM dose deformation | |
| L3 | endogenous rhythm + exogenous drug | rhythm survives superposition | Dev = 1.410 (Endo) = 1.410 (Mix); Δt = 24.00 h | weak baseline signal read through the drug |
| L4 | 15 real MPA profiles | gut-microbiome anchor | ρ = 0.815 (TPK) vs 0.754 (manual EHR%) vs ρ ≤ 0.38 (NCA exposure) | value on real, sparse clinical data |

## 2. Theoretical Framework

### 2.1 Pharmacokinetic systems as dynamical flows

The ADME system involves a multitude of interacting entities—drug molecules in plasma, tissue-bound drug, enzyme occupancy states, transporter activities, metabolite concentrations, and physiological rhythms—defining a high-dimensional state vector $x(t)$ evolving on a smooth, low-dimensional manifold $M$, constrained by mass balance and physiological regulation:

$$\dot{x}(t) = f(x(t)) + u(t), C(t) = h(x(t)) \tag{2.1}$$

where $f$ encodes the intrinsic pharmacokinetic dynamics, $u(t)$ the externally imposed dosing regimen, and $h: M \to \mathbb{R}$ the observation function that projects the high-dimensional trajectory onto a single dimension. The manifold $M$ is the set of states the system actually visits; its dimension reflects the true degrees of freedom. Information is lost in this projection: features clearly separated on $M$ may overlap in the scalar record, and geometric structures such as loops may be flattened beyond recognition. The central hypothesis of topological pharmacokinetics (TPK) is that despite this loss, the concentration–time series retains sufficient structure to recover essential topological features of $M$—guaranteed under generic conditions by Takens' embedding theorem [24–26]. TPK is accordingly not a mechanism-identification engine: it summarizes shape constraints imposed by the data, complementing—never replacing—model-based inference.

### 2.2 Normalization and delay embedding

Prior to constructing delay vectors, the series is normalized by its maximum to eliminate dose-magnitude confounding and ensure scale invariance of all subsequent features:

$$\widehat{C}(t) = C(t)/C_{max}, C_{max} = max_t C(t) \tag{2.2}$$

Delay vectors are then constructed from the normalized series:

$$v_i = \left(\widehat{C}(t_i), \widehat{C}(t_i + \tau)\right) \tag{2.3}$$

where $t_i$ is the $i$-th sampling time and $\tau$ the time delay; the set $P = \{v_i\}$ constitutes the reconstructed point cloud (embedding dimension $m = 2$ throughout). Classical embedding-parameter criteria—mutual information for the delay time [27] and false-nearest-neighbors for the embedding dimension [28]—presuppose long, densely sampled, stationary attractor data [29]; clinical concentration–time series are short, transient, and nonstationary, so $m$ and $\tau$ are fixed a priori from the simulation studies and their sensitivity reported explicitly (Section 3.4). Before homology is computed, the embedded trajectory is standardized by re-sampling at equally spaced arc-length positions (a fixed 400 points), so that the scale of the filtration complex is independent of the original sampling density. A single delay coordinate suffices to unfold prominent cyclic and curved structures of concentration–time trajectories while permitting direct visualization. Takens' theorem guarantees preservation of the manifold's topological features under genericity conditions on $h$ and $f$; we note explicitly that irregular, sparse clinical sampling does not satisfy the theorem's hypotheses, and we therefore treat embedding-based features as shape statistics of the observed curve rather than as reconstructions of an underlying attractor (Section 2.4; Discussion).

Two strategies address sparse data. (i) *Interpolation-based densification*: profiles sampled at 5–16 irregular time points are interpolated onto a uniform grid by a log-up/linear-down (LULD) rule—log-linear across descending segments, linear across ascending segments, no smoothing—and the embedding is applied to the interpolated curve. Interpolation affects feature computation only; all conventional metrics are computed from the raw samples. (ii) *Model-assisted probabilistic embedding* (framework component, not exercised in the present applications): a population PK model is fit to the full dataset; for a sparsely observed subject, Bayesian posterior estimation yields a predictive distribution over plausible dense trajectories, an ensemble of $K$ such trajectories is simulated, and topological stability is tested via pairwise bottleneck distances $D$ of their persistence diagrams. If the maximum bottleneck distance falls below a stability threshold (the 95th percentile of null distances from permuted data), the topology is deemed robust to posterior uncertainty; otherwise the method transparently reports insufficiency. Crucially, the population model serves as a principled interpolation engine, not an interpretive lens: the topology is interrogated after densification, not pre-judged by it.

## 2.3 Persistent homology

Persistent homology tracks topological features across a multiscale Vietoris–Rips filtration of the embedded point cloud $P$ [11,12].

**H0 (connected components).** For a one-dimensional concentration–time curve, zero-dimensional persistence is equivalent to a prominence analysis of the time series: the prominence of a peak—its vertical distance to the highest contour separating it from a higher peak—equals exactly the persistence of the corresponding connected component when the filtration parameter is interpreted as the concentration level (a superlevel-set construction). Each local maximum of the interpolated curve with prominence $\geq 5\%$ of $C_{max}$ thus generates one H0 bar whose length equals that prominence; in this prominence-based construction every bar is finite—the dominant peak's bar terminates at the minimum of the observed series—and no essential (infinite) bar is recorded. H0 features encode the hierarchy of peak-like structure: a secondary peak well separated from its neighbors generates a persistent H0 bar.

**H1 (loops).** One-dimensional persistence is computed on the normalized two-dimensional delay vectors. A loop is significant if its persistence exceeds a fixed threshold $\varepsilon_{noise} = 0.015$ in normalized units—an empirical cutoff validated on simulated linear systems (Section 3.1), separating genuine cyclic structures from noise under dense sampling; for substantially different signal-to-noise regimes, a permutation-based threshold is recommended (Supplementary Text S2). In pharmacokinetic terms, H1 generators are the signature of cyclic or strongly curved processes: enterohepatic recirculation, where a secondary peak creates a closed excursion in delay-embedded space; saturable elimination, where the transition between kinetic orders produces a bowed trajectory; and endogenous rhythms, where a periodic physiological driver imprints a stable cycle.

**Stability.** Persistence diagrams are stable under perturbations of the input in the bottleneck and Wasserstein metrics [14,15]: small changes in the measured curve induce correspondingly bounded changes in the diagram. This guarantee applies to persistence diagrams; the behavior of derived descriptors also depends on normalization, thresholding, and vectorization. It therefore motivates, rather than replaces, the empirical noise-robustness analyses in Section 3.4. Persistent entropy is additionally stable in a distributional sense [30].

All persistent homology is computed with ripser [31], built on the standard persistent-homology computation framework [32]; peak prominence and curvature with SciPy [33].

## 2.4 The pharmacokinetic topological invariant (PTI) vector and extended descriptors

From the point cloud $P$ and its persistence diagrams $PD_0(P)$, $PD_1(P)$ we extract the **pharmacokinetic topological invariant (PTI) vector**—nine components, organized into two classes following barcode-vectorization strategies such as persistence landscapes [34] (Table 2). Class A components are strict topological invariants; Class B are geometry-infused descriptors computed after topological structure is established. Throughout the results, tables, and figures, components are referred to by their descriptive names:

$N_{sig} \equiv \beta_1$, Pers_Ratio $\equiv P_{ratio}$, Entropy_PE $\equiv E_{PE}$, Integral_PI $\equiv \Pi_{int}$, Dev $\equiv D_{dev}$, Pers_main $\equiv P_{main}$, and Decouple $\equiv R_{saddle}$; $N_{H_0}$, $N_{PTP}$, $\Lambda_{NL}$, and $\Delta t$ carry over unchanged.

**Table 2.** The PTI vector: nine core descriptors (Class A strict topological invariants; Class B geometry-infused) and three simulation-level extensions—symbol, definition, and the shape information each captures.

| Component (symbol / name) | Class | Definition | Shape information captured |
|---|---|---|---|
| $\beta_1$ / N_sig | A | significant H1 loop count, Eq (2.4) | recurrent, loop-like dynamics (recirculation, rhythms) |
| $N_{H_0}$ | A | H0 bars surviving the 5%·$C_{max}$ prominence threshold | extent of multi-peak organization |
| $P_{main}$ / Pers_main | A | dominant H0 bar persistence (primary-peak prominence, $C_{max}$-normalized) | prominence of the primary peak |
| $P_{ratio}$ / Pers_Ratio | A | secondary-persistence ratio, Eq (2.5) | relative prominence of secondary structure |
| $\Pi_{int}$ / Integral_PI | A | $\sum_i prom_i/C_{max}$ | total multi-scale structure |
| $E_{PE}$ / Entropy_PE | A | persistent entropy, Eq (2.6); $2^{E_{PE}}$ = effective number of persistent peaks | heterogeneity of the persistence spectrum |
| $N_{PTP}$ | B | curvature-peak count along the embedded trajectory, Eq (2.7) | fine-scale structure (noise-sensitive) |
| $D_{dev}$ / Dev | B | terminal-segment chord deviation, Section 2.4 | departure from log-linear terminal decline |
| $\Lambda_{NL}$ | B | maximum finite H1 persistence / Diam($P$), Eq (2.8) | nonlinear-type deviation |
| $R_{saddle}$ / Decouple | ext. | $C_{valley}/C_{max}$ | separation of two kinetic events |
| $\Delta t$ | ext. | first-two-peak interval | cycle-period constraint |
| $P_{max}(H_1)$ | ext. | $max\{d_i - b_i\}$ | strongest loop, absolute scale |

*Class A.* The loop count

$$\beta_1 = |\{(b_i, d_i) \in PD_1(P) \mid d_i - b_i > \varepsilon_{noise}\}| \tag{2.4}$$

counts significant H1 generators (denoted $N_{sig}(H_1)$ in the preprint version of this framework). The H0 feature count $N_{H_0}$ counts bars surviving the prominence threshold. The main persistence $P_{main}$ is the persistence of the dominant H0 bar (the primary peak's prominence, $C_{max}$-normalized), and the secondary persistence ratio

$$P_{ratio} = \frac{P_{H_0}^{(2)}}{P_{H_0}^{(1)}} \tag{2.5}$$

is the persistence of the second-most-persistent H0 bar relative to the dominant bar (0 if no second bar survives; → 1 for co-dominant peaks). The integrated persistence $\Pi_{int} = \sum_i prom_i/C_{max}$ sums the prominences of all threshold-surviving bars, and the persistent entropy

$$E_{PE} = -\sum_i p_i log_2 p_i, p_i = prom_i/\sum_j prom_j \tag{2.6}$$

summarizes the heterogeneity of the persistence spectrum ($2^{E_{PE}}$ = effective number of persistent peaks).

*Class B.* The number of phase-transition points $N_{PTP}$ locates curvature peaks along the embedded trajectory: the local curvature

$$\kappa(t) = \frac{|\dot{x}\ddot{y} - \dot{y}\ddot{x}|}{(\dot{x}^2 + \dot{y}^2)^{3/2}} \tag{2.7}$$

of the delay trajectory $\gamma(t) = (C(t), C(t+\tau))$ is computed by central finite differences on the unnormalized coordinates (the relative prominence threshold below makes the count invariant to amplitude scaling), and peaks in $\kappa(t)$ with prominence $\geq max(10^{-5}, 0.05 max_t \kappa)$—with the three edge points at each end trimmed—are counted. The deviation statistic $D_{dev}$ is the maximum perpendicular deviation of the delay-embedded terminal post-peak segment from its endpoint chord, normalized by the $x$-span (concentration drop) of that segment; for first-order (mono-exponential) decay the embedded trajectory is a straight line, so $D_{dev}$ quantifies departure from log-linear terminal decline. The nonlinearity index

$$\Lambda_{NL} = \frac{max\{d_i - b_i\}}{Diam(P)} \quad (2.8)$$

is the maximum finite H1 persistence divided by the diameter of the point cloud, rendering it comparable across dose levels and individuals.

*Simulation-level extensions.* Three further descriptors are reported in the simulation studies: the saddle decoupling index $R_{saddle} = C_{valley}/C_{max}$, reported as *Decouple* in the results ($C_{valley}$ = minimum concentration between the first two detected peaks; → 0 cleanly decoupled, → 1 merged); the inter-peak interval $\Delta t$ (time difference between the first two detected peaks, in chronological order; approximates the cycle period when a significant H1 loop is present); and the maximum H1 persistence $P_{max}(H_1) = max\{d_i - b_i\}$ (zero for an empty diagram).

Table 2 maps each component to the shape information captured; full degenerate-case handling and normalizations follow Table 2 and Supplementary Text S1.

### 2.5 Statistical inference

All analyses are exploratory. Associations between descriptors and external variables are assessed with Spearman rank correlations, confirmed where headline claims are at stake by permutation tests ($10^5$ permutations), with 95% percentile bootstrap confidence intervals ($10^4$ paired resamples) as precision measures; rank-partial correlations control for specified covariates. Between-group differences use Kruskal–Wallis tests with pairwise Mann–Whitney U follow-ups. Multiplicity is controlled within descriptor families by Benjamini–Hochberg FDR, with Bonferroni-corrected significance reported alongside. Embedding parameters ($\tau$, prominence and persistence thresholds, interpolation density) are fixed a priori from the simulation studies rather than re-optimized per application; their sensitivity is reported as a grid wherever an association is claimed (Section 3.4). Formal inference on persistence diagrams themselves (e.g., permutation tests on persistence landscapes) was not pursued given the sample sizes involved; rank-based descriptor-level tests are used throughout.

## 3. Simulation Studies

The simulation program is organized as a three-layer progression with ground truth, followed by a systematic operating-characteristics analysis (Section 3.4). Layer 1 establishes the negative control: linear pharmacokinetic systems must not generate spurious cyclic signals. Layer 2 asks whether nonlinearity—the most common departure from linear kinetics—is resolved as shape deformation. Layer 3 asks whether a weak endogenous rhythm survives superposition with a large exogenous drug signal. The baseline simulations are dense and noise-free by design (300 points over 24 h in Layers 1–2; 1,201 points over 120 h in Layer 3); noise, sampling sparsity, and embedding-parameter sensitivity are studied systematically in Section 3.4, where they belong. All features are computed with the fixed pipeline of Section 2 (τ = 5 samples, prominence threshold 5% of $C_{max}$, H1 persistence threshold $\varepsilon_{noise}$ = 0.015); no parameter is re-tuned per system.

### 3.1 Layer 1: Linear benchmark — monophasic controls and double-peak signatures

**Design.** Six linear systems spanning the standard pharmacokinetic families: IV bolus with one and two compartments; oral administration with one and two compartments; a double-site absorption (DSA) model in which a lagged second absorption site (lag 4 h, fraction 0.5) partially overlaps the first; and an enterohepatic recirculation (EHC) model in which a gallbladder compartment empties at 12 h, returning drug to the central compartment. DSA and EHC both produce two-peaked concentration–time profiles—but by different mechanisms: coupled parallel absorption (DSA) versus sequential physiological cycling (EHC). The question is whether TPK's diagnostic hierarchy separates them.

**Results (Table 3, Fig 2).** All four monophasic linear systems returned N_sig(H1) = 0: no spurious loops in these negative controls. Multi-compartment and absorption curvature was visible where it exists (N_PTP ≤ 1, D_dev ≤ 0.032) but never closed into a cycle. Oral DSA returned two significant H1 generators (N_sig = 2, Entropy_PE = 0.553, Δt = 3.53 h, Decouple = 0.470): the topological signature of two partially coupled absorption events. Oral EHC returned one H1 generator with a substantially longer inter-peak interval (Δt = 12.20 h) and near-complete decoupling (Decouple = 0.157)—almost four times the DSA interval, consistent with the physiological time scale of biliary excretion and gallbladder emptying.

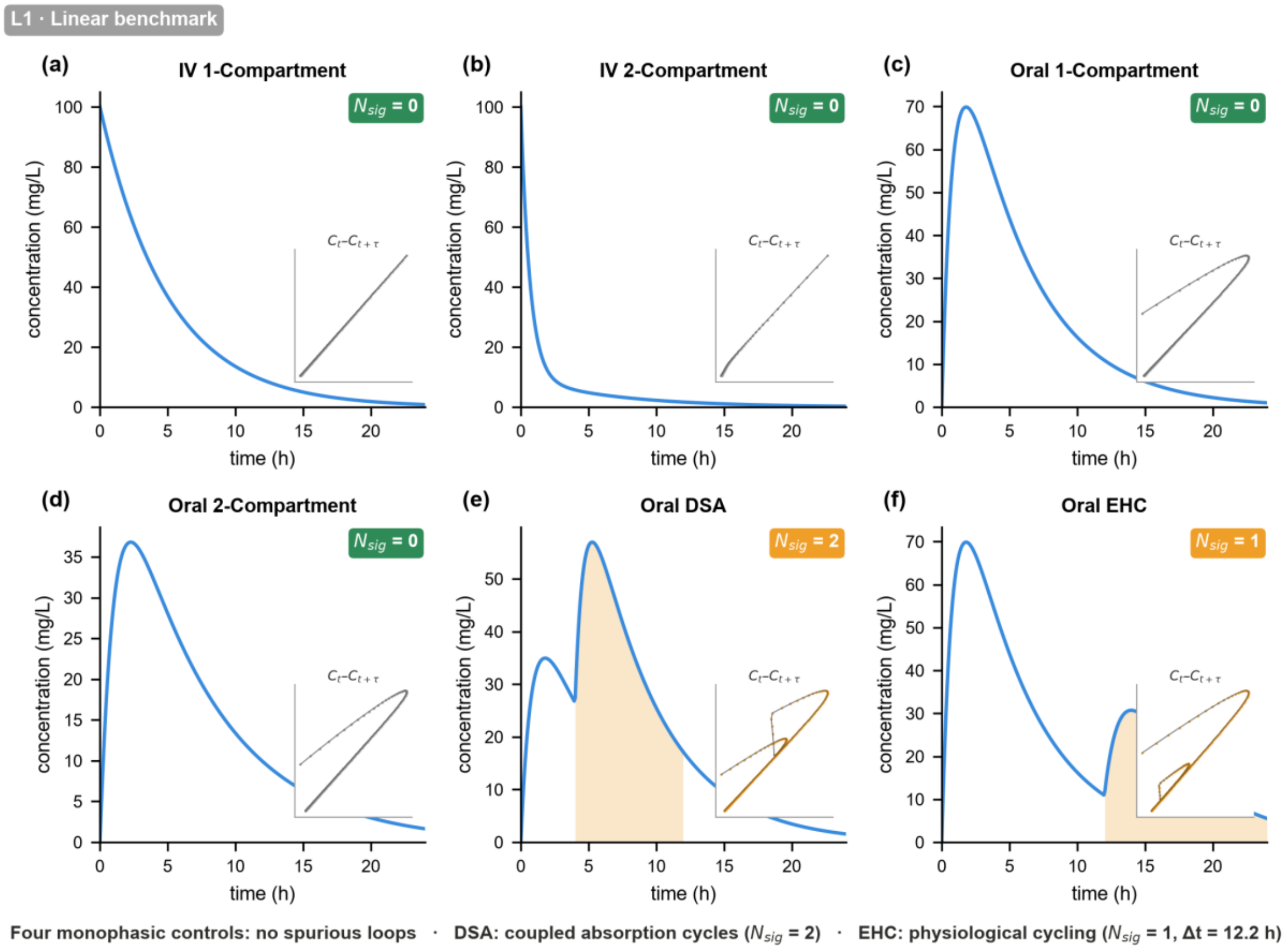

**Fig 2. Layer 1 (linear benchmark): no spurious loops in four monophasic controls—while DSA and EHC show distinct loop signatures.** Six systems in a 2×3 grid (IV 1- and 2-compartment; oral 1- and 2-compartment; double-site absorption, DSA; enterohepatic recirculation, EHC). Each panel: concentration–time curve with secondary peaks shaded amber, delay-embedded trajectory inset. Green badges: N_sig = 0 (linear systems—no spurious loops); amber badges: N_sig = 2 (DSA) and 1 (EHC). DSA and EHC both produce two peaks, but are separated by decoupling (0.47 vs 0.16) and inter-peak interval (3.53 h vs 12.20 h)—coupled parallel absorption vs sequential physiological cycling.

**Table 3.** Layer 1: the six linear benchmark systems—full descriptor values (N_sig, N_PTP, D_dev, Entropy_PE, Pers_Ratio, Δt, Decouple).

| System | N_sig | N_PTP | D_dev | Entropy_PE | Pers_Ratio | Δt (h) | Decouple |
|---|---|---|---|---|---|---|---|
| IV 1-compartment | 0 | 0 | 0.0000 | -0.000 | 0.000 | 0.00 | 0.000 |
| IV 2-compartment | 0 | 1 | 0.0206 | -0.000 | 0.000 | 0.00 | 0.000 |
| Oral 1-compartment | 0 | 1 | 0.0318 | -0.000 | 0.000 | 0.00 | 0.000 |
| Oral 2-compartment | 0 | 1 | 0.0266 | -0.000 | 0.000 | 0.00 | 0.000 |
| Oral DSA | 2 | 3 | 0.0292 | 0.553 | 0.147 | 3.53 | 0.470 |
| Oral EHC | 1 | 4 | 0.0342 | 0.786 | 0.307 | 12.20 | 0.157 |

**Interpretation.** The DSA–EHC contrast establishes the diagnostic hierarchy in a single comparison. At the first level, D_dev and N_PTP register ordinary curvature without asserting cyclicity. At the second, N_sig ≥ 1 confirms genuine cyclic structure—present only in DSA and EHC among the six systems. At the third, the Decouple index separates sequential physiological cycling (EHC, 0.16) from coupled parallel processes (DSA, 0.47). At the fourth, Δt provides mechanistic constraints: short Δt suggests parallel absorption; long Δt suggests sequential cycling. No single conventional parameter provides a comparable stratified diagnosis of a double-peaked profile.

### 3.2 Layer 2: Nonlinear deformation — 24 scenarios

**Design.** A full factorial of 8 structural configurations (IV/oral × 1/2 compartments × linear/Michaelis–Menten elimination, $V_{max}$ = 10, $K_m$ = 5) × 3 dose levels (200, 1,000, 4,000 mg), i.e. 24 scenarios. MM elimination is the canonical model-free target: saturation straightens the trajectory at high concentration; the return to first-order kinetics as concentration falls through $K_m$ bends it again. Because each nonlinear scenario has a linear counterpart at the same dose, shape invariants can be compared directly.

**Results (Table 4, Fig 3).** Across all 24 scenarios, N_sig(H1) = 0. This is not a detection failure but a precise topological distinction: nonlinear elimination produces *curvature without cycles*—in contrast to the genuine H1 generators of DSA and EHC (Layer 1) and of endogenous rhythms (Layer 3). In these benchmark systems, the H1 channel thus distinguished the tested cyclic processes from curvature produced by nonlinear elimination.

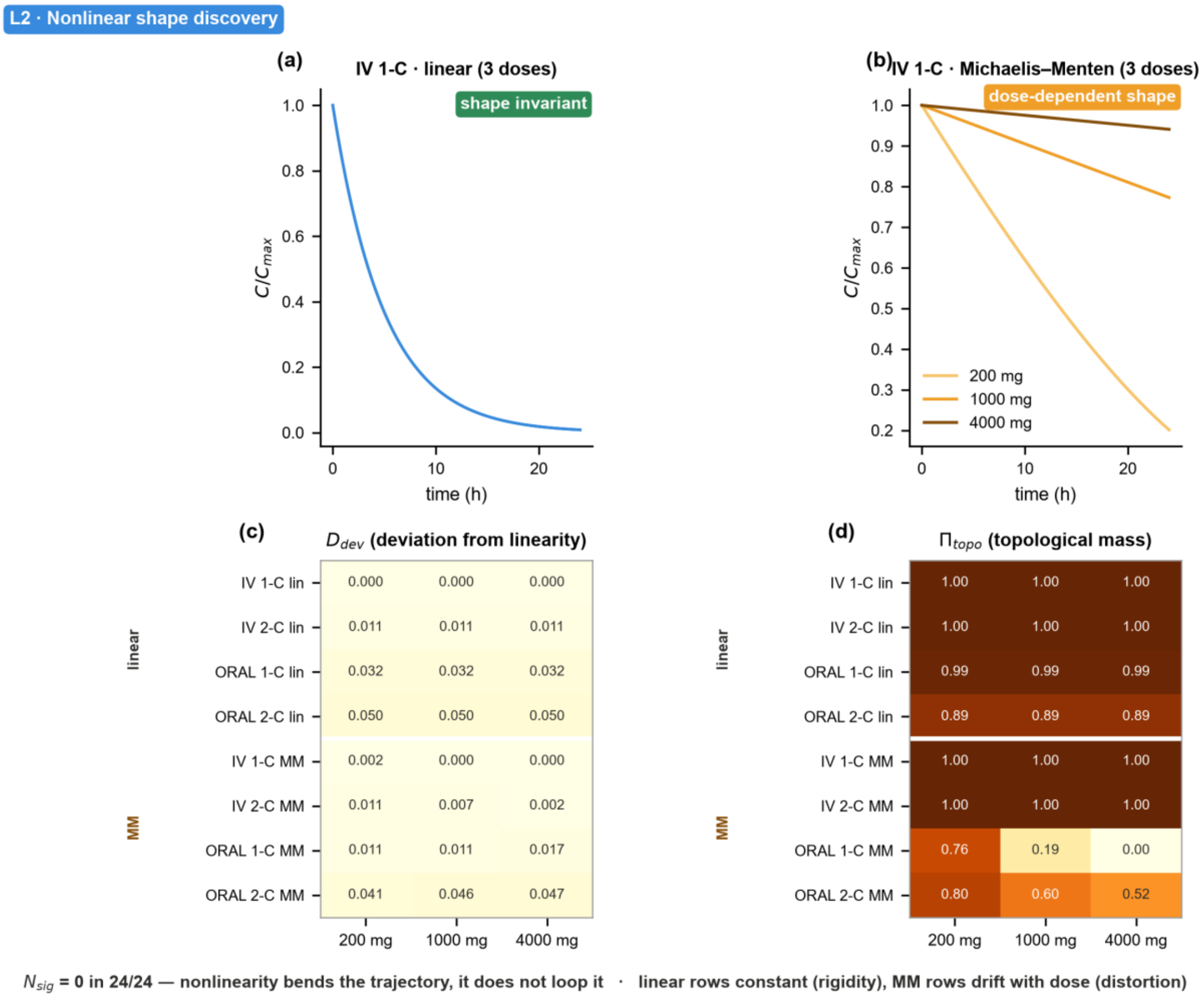

**Fig 3. Layer 2 (nonlinear deformation): linear systems are shape-rigid; Michaelis–Menten elimination deforms the shape with dose—curvature without cycles in all 24 scenarios.** (a) IV 1-compartment, linear elimination, three doses normalized to $C_{max}$: the trajectories superimpose exactly (shape rigidity; green badge). (b) The same system with Michaelis–Menten elimination: the three normalized trajectories separate—higher doses plateau longer (dose-dependent deformation). (c, d) Descriptor heat maps, 8 structural configurations × 3 doses: D_dev and Integral_PI are constant across doses in every linear row (rigidity) and drift in every MM row (deformation); oral MM Integral_PI collapses 0.76 → 0.19 → 0.00. N_sig = 0 in 24/24 scenarios—nonlinearity bends the trajectory; it does not loop it.

**Table 4.** Layer 2: 24-scenario summary—shape rigidity of linear systems (descriptor invariance across doses) vs. dose-dependent deformation of Michaelis–Menten systems.

| Configuration | D_dev 200 mg | 1000 mg | 4000 mg | Integral_PI 200 mg | 1000 mg | 4000 mg | N_sig |
|---|---|---|---|---|---|---|---|
| IV 1-C linear | 0.000 | 0.000 | 0.000 | 1.000 | 1.000 | 1.000 | 0 |
| IV 2-C linear | 0.011 | 0.011 | 0.011 | 1.000 | 1.000 | 1.000 | 0 |
| ORAL 1-C linear | 0.032 | 0.032 | 0.032 | 0.986 | 0.986 | 0.986 | 0 |
| ORAL 2-C linear | 0.050 | 0.050 | 0.050 | 0.895 | 0.895 | 0.895 | 0 |
| IV 1-C MM | 0.002 | 0.000 | 0.000 | 1.000 | 1.000 | 1.000 | 0 |
| IV 2-C MM | 0.011 | 0.007 | 0.002 | 1.000 | 1.000 | 1.000 | 0 |
| ORAL 1-C MM | 0.011 | 0.011 | 0.017 | 0.759 | 0.189 | 0.000 | 0 |
| ORAL 2-C MM | 0.041 | 0.046 | 0.047 | 0.800 | 0.604 | 0.518 | 0 |

Linear systems exhibited shape rigidity: with all descriptors computed on the $C_{max}$-normalized curve, D_dev and Integral_PI were numerically identical across the three dose levels (e.g. oral 1-compartment: Integral_PI = 0.986, 0.986, 0.986). In contrast, every MM system exhibited dose-dependent deformation in at least one descriptor. The most robust indicator was Integral_PI in oral MM systems, which collapsed monotonically with dose (oral 1-compartment: 0.759 → 0.189 → 0.000 at 200/1,000/4,000 mg): saturable elimination prolongs exposure and progressively flattens the normalized attractor's peak hierarchy. In IV MM systems, D_dev *decreased* with dose—saturation straightens the trajectory, so the high-dose curve is closer to a straight line than the curved first-order reference.

**Interpretation.** Two independent shape signatures—shape rigidity of linear systems (invariance under dose scaling) and dose-dependent deformation of MM systems—separate linear from nonlinear kinetics without fitting any model and without knowing $K_m$. We note explicitly that the preprint version of this work [35] additionally reported a "N_PTP + 1 rule" (each MM system exhibiting exactly one more curvature transition point than its linear counterpart); that rule was not reproducible under the present pipeline over 24-h and 48-h windows and is not claimed here. The rigidity/deformation contrast is the stronger evidence, being exact under the present implementation.

## 3.3 Layer 3: Endogenous rhythm beneath an exogenous drug

**Design.** Three signals over 120 h (0.1-h resolution): an endogenous 24-h rhythm C(t) = 10 + 4·cos(2πt/24) (amplitude 4% of the exogenous peak); an exogenous one-compartment IV bolus ($C_{max}$ = 100, dosed at t = 12 h, k = 0.2 $h^{-1}$); and their linear superposition. The exogenous signal is analyzed on the post-dose window (t ≥ 12 h): the pre-dose zero segment causes the embedded point cloud to pile up at the origin, producing a spurious loop that is an artifact of the window, not of the pharmacokinetics. The contrastive strategy compares the PTI vectors of the pure and mixed signals rather than attempting physical separation of the components.

**Results (Table 5, Fig 4).** The mixed signal inherited the rhythm's H1 generator: N_sig = 1 for Endo and Mix, 0 for Exo. Dev was the most striking finding: Endo = 1.410, Exo = 0, Mix = 1.410—identical to three decimal places. Dev is thus a rhythm-specific shape statistic, invariant under superposition with a linear exogenous drug: a non-zero Dev in an expected-linear system is a diagnostic flag for an unmodeled periodic baseline, requiring no baseline measurement, no pre-dose sampling, and no parametric model of the rhythm. The detected inter-peak interval Δt = 24.00 h for the pure rhythm equals its true period exactly.

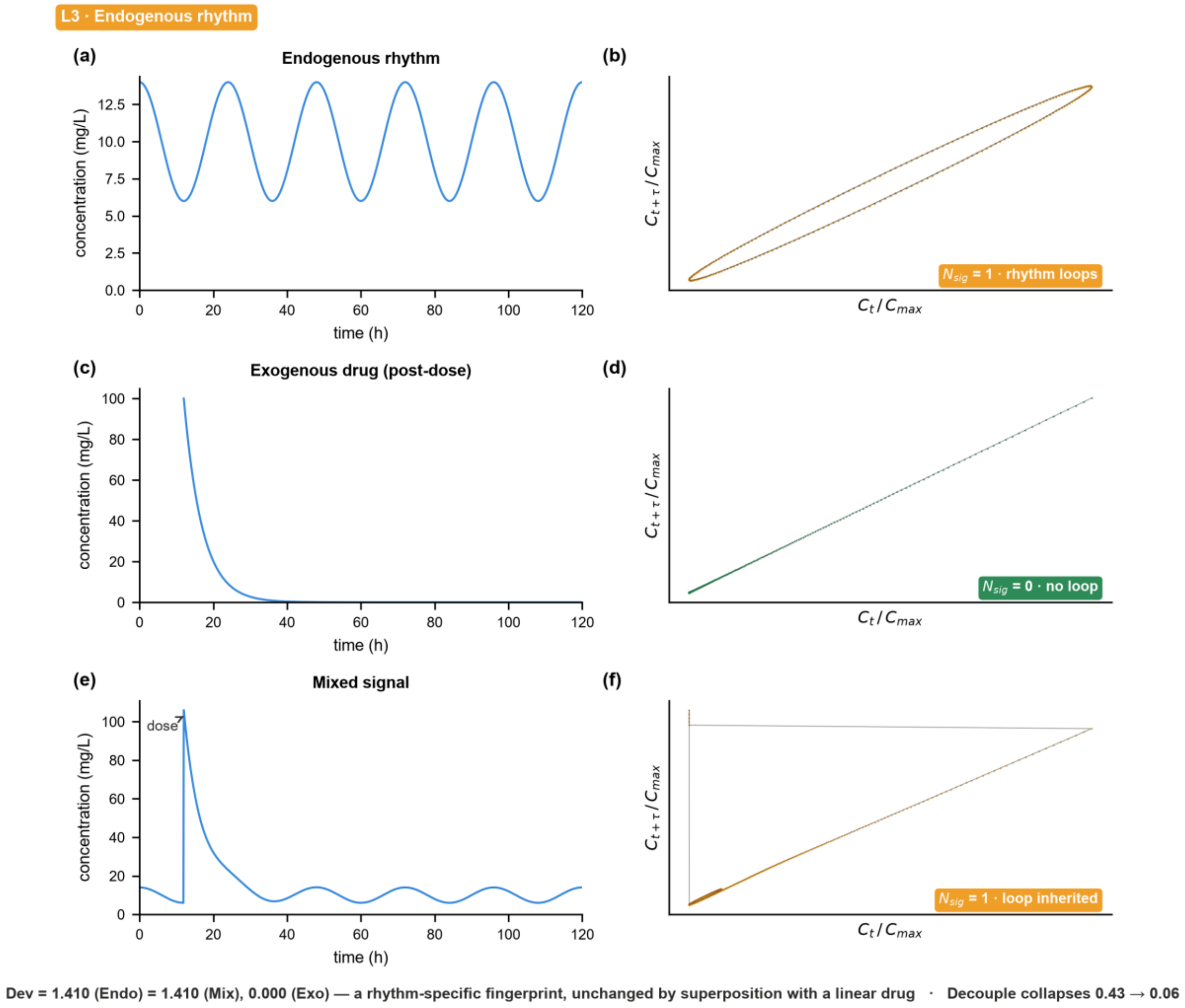

**Fig 4. Layer 3 (endogenous rhythm): the rhythm's loop and shape distortion survive superposition with a large exogenous drug signal.** Three rows—endogenous rhythm (C = 10 + 4·cos(2πt/24)), exogenous drug (IV bolus at 12 h), and their mixture; left column: concentration–time curves (dosing arrow); right column: delay-embedded point clouds. The rhythm loops (N_sig = 1, badge); the drug does not (N_sig = 0, analyzed on the post-dose window); the mixture inherits the loop (N_sig = 1). Bottom: Dev = 1.410 (Endo) = 1.410 (Mix) vs 0 (Exo)—the rhythm's shape distortion passes through superposition unchanged—and the detected inter-peak interval equals the true 24-h period exactly.

**Table 5.** Layer 3: the three signals (endogenous / exogenous / mixed)—N_sig inheritance, Dev specificity, Decouple collapse.

| Signal | N_sig | D_dev | Entropy_PE | Δt (h) | Decouple |
|---|---|---|---|---|---|
| Endogenous rhythm | 1 | 1.410 | 2.000 | 24.00 | 0.429 |
| Exogenous drug (post-dose window) | 0 | 0.000 | -0.000 | 0.00 | 0.000 |
| Mixed signal | 1 | 1.410 | 0.998 | 35.90 | 0.064 |

The Decouple index collapsed from 0.429 (Endo) to 0.064 (Mix)—near-complete coupling. In the pure endogenous state the rhythm's peaks and troughs are distinct, well-separated excursions; under the drug's steep exponential decay they become tightly bound to the elimination trajectory. This *Decouple Collapse* is a pharmacodynamic signature readable from PK data alone: the drug does not merely coexist with the rhythm—it alters the rhythm's dynamical relationship to the observed system.

**Interpretation.** Three independent signatures emerge from the contrastive design: N_sig inheritance, Dev specificity, and Decouple collapse. Their combination completes the diagnostic hierarchy across layers: nonlinear elimination *bends* the trajectory (curvature without cycles, N_sig = 0); endogenous rhythms *loop* it (cycles, N_sig = 1, inherited by the mixture); physiological recirculation *loops* it with a drug-scaled inter-peak interval (Layer 1).

### 3.4 Operating characteristics: noise, sampling density, and embedding parameters

A recurring objection to shape-sensitive methods is fragility: perhaps the descriptors merely register noise or sampling artifacts. We quantify the operating envelope of the pipeline on two carriers—the EHC system of Layer 1 (a drug-scale cycle) and the endogenous rhythm of Layer 3 (a baseline-scale cycle, analyzed as the mixed signal would see it: weak relative to a co-administered drug)—across three stress axes. Each condition is replicated over 20 independent noise realizations (Fig 5; full grids in Supplementary Table S3).

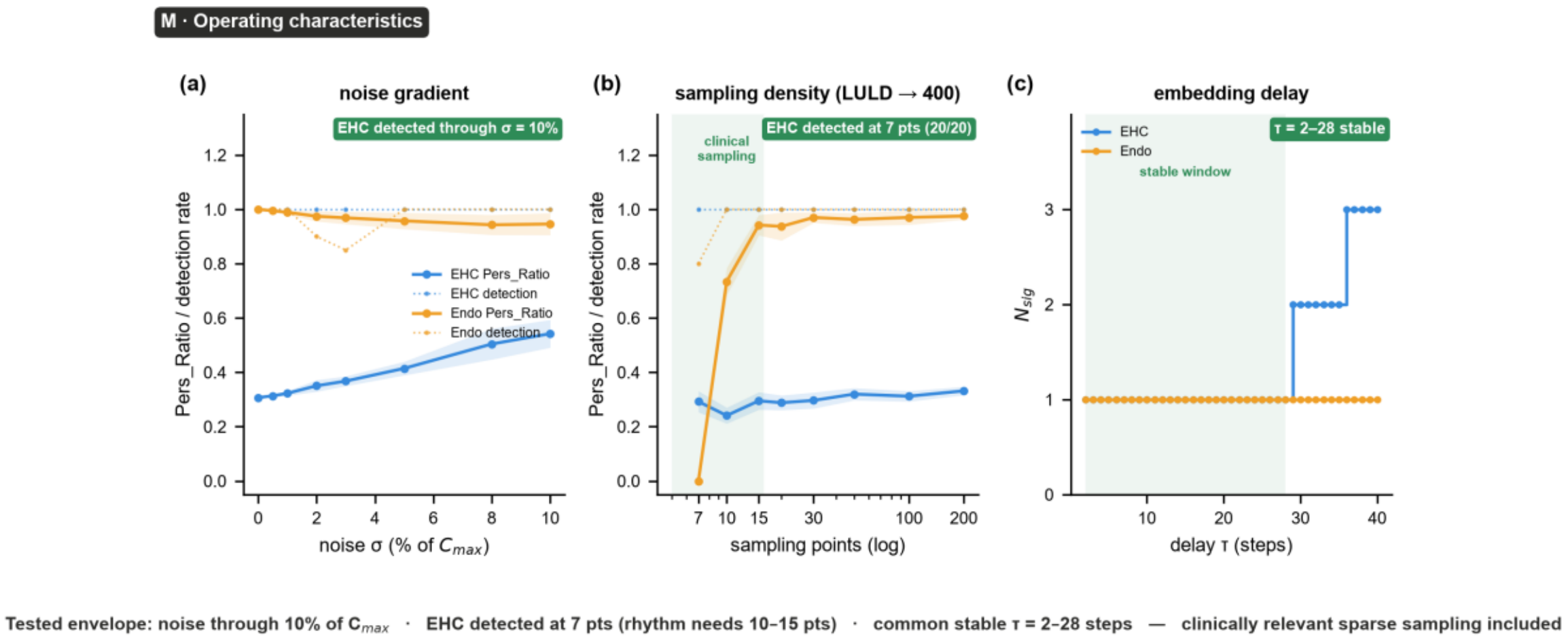

**Fig 5. Operating characteristics: the tested envelope includes clinically relevant sparse-sampling conditions.** (a) Additive noise, σ = 0–10% of $C_{max}$, 20 realizations per condition: the EHC cycle is detected in 20/20 realizations at every noise level (green badge); Pers_Ratio drifts upward at high noise, as noise adds shallow prominence. (b) Sampling density, thinned from dense grids (300 points for EHC; 1,201 for endogenous rhythm) to 7 points and re-interpolated (LULD): the EHC cycle is detected in 20/20 realizations at every density down to 7 points (Pers_Ratio 0.29 vs 0.31 dense); the endogenous rhythm requires 10–15 points. The green band marks the clinical sampling regime (5–16 points)—the MPA application operates inside the envelope. (c) Embedding delay τ, 2–40 grid steps: N_sig stable (window shaded green) for τ = 2–28, with predictable count inflation beyond—excessive delay over-unfolds the single excursion into multiple apparent cycles.

**Measurement noise.** Additive Gaussian noise, σ from 0 to 10% of $C_{max}$. Loop detection was retained across the tested gradient: the EHC cycle was detected in 20/20 realizations at every σ, and the endogenous cycle in ≥ 17/20 (85%) at every σ. For orientation, 20/20 detections give an exact Clopper–Pearson 95% CI of 83.2–100%, and 17/20 gives 62.1–96.8%; these small-sample intervals do not justify a claim of certainty. Pers_Ratio drifted upward at high noise (EHC: 0.31 → 0.54 from σ = 0 to 10%), as expected since noise adds shallow prominence to the persistence spectrum. Detection of the *presence* of a cycle is far more robust than its precise magnitude.

**Sampling density.** Uniform thinning from the dense grids (300 points for EHC; 1,201 for endogenous rhythm) down to 7, 10, 15, 20, 30, 50, 100, 200 points, each thinned profile re-interpolated by the LULD rule of Section 2.2. The EHC cycle was detected in 20/20 realizations at *every* density, including 7 points (Pers_Ratio 0.29 vs 0.31 dense), because interpolation-based densification preserves the two-peak structure as long as the sampling times resolve both peaks. The endogenous rhythm behaved differently: at 7 points detection collapsed (80% realizations, Pers_Ratio 0.00) and recovered at 10–15 points (Pers_Ratio 0.73 → 0.94). A cycle whose period is comparable to the observation window requires sampling that resolves the oscillation itself, not merely its envelope. This asymmetry defines the practical envelope: **drug-scale cycles (secondary peaks) tolerate extreme sparsity; rhythm-scale cycles require ≥ 10–15 points.** The clinical sampling of Section 4 (5–16 points per profile, targeting peak and trough structure) falls within the drug-cycle envelope—precisely the regime in which the MPA application operates.

**Embedding parameter τ.** N_sig was stable for τ = 2–28 samples (0.2–2.8 h at the simulated resolution) on both carriers, with the EHC cycle's count rising predictably to 2 at τ = 29–35 and 3 at τ ≥ 36: excessive delay over-unfolds the trajectory and splits the single excursion into multiple apparent cycles, the expected qualitative behavior from embedding theory. The endogenous cycle was stable (N_sig = 1) across the entire τ = 2–40 grid. All results reported in Sections 3.1–3.3 use τ = 5, well inside the stable window.

**Summary.** The tested operating envelope includes clinically relevant sparse-sampling conditions: additive noise up to at least 10% of $C_{max}$; sparse sampling down to 7 points for peak-structure cycles; and a generous, non-critical choice of τ. At the tested extremes, degradation was interpretable (detection loss at extreme sparsity; count inflation at extreme delay), but these two carrier systems do not establish universal behavior outside the tested envelope. The envelope itself is reported so that any future application can verify its sampling regime falls within it—as the MPA data of Section 4 does.

## 4. Clinical Application: Mycophenolic Acid Enterohepatic Recirculation

The simulation layers establish what the pipeline *can* do under ground truth; the clinical layer asks what it *does* do on real, sparse, noisy, human data—in a pharmacological setting where shape is the mechanism. Mycophenolic acid (MPA) undergoes pronounced enterohepatic recirculation (EHR) [36]: glucuronidated MPA is excreted into bile, deconjugated by gut-microbial β-glucuronidase, and reabsorbed, producing a characteristic secondary concentration peak whose timing and magnitude vary across patients. Quantifying that recirculation from routine concentration–time data is a concrete clinical problem: MPA dosing in transplant recipients is exposure-guided, EHR is a recognized source of between-patient variability [36], and the available quantifications—trough-anchored post-peak AUC ratios, scalar rebound heights, or manually keyed EHR fractions as in the source study [10]—share two structural limitations: the trough or onset sample is chosen by the operator, and every metric is single-scale, scoring a gradual shoulder and an isolated secondary peak of equal rebound height alike. The reabsorption pathway passes directly through the gut microbiome—so if the shape of the concentration–time curve carries mechanistic information, the microbiome should leave a fingerprint on it. This is precisely the kind of signal that scalar exposure metrics (AUC, $C_{12}$) average away.

### 4.1 Data

We analyzed all 15 individual MPA plasma concentration–time profiles reported in Drevland et al. [10] (Microbiome, 2025; CC BY 4.0), digitized from the published supplementary figure with WebPlotDigitizer [37] (tick-calibrated axes; extraction fidelity verified by dual-channel quality control, Supplementary Tables S1 and S4). This is a secondary analysis of openly licensed published data; no new human-subject data were collected, and ethical approval rests with the source study. The source study performed 12-h PK investigations in 10 kidney transplant recipients 3–6 weeks post-transplantation, five of whom repeated the investigation at one year (15 profiles; patient–profile pairing is not recoverable from the publication). All patients received mycophenolate mofetil 750 mg twice daily at steady state, with 16 samples per profile (238 concentration–time pairs; two profiles contributed 15). No smoothing was applied; profiles were densified by LULD interpolation (Section 2.2) only for topological feature computation and plotting, with all conventional metrics computed from the raw samples.

Conventional comparator metrics were computed independently of the topological pipeline: NCA summaries ($C_{max}$, $T_{max}$, $AUC_{0-12}$, predose $C_0$) and four operationalizations of recirculation magnitude—the algorithmic EHC AUC fraction (post-trough AUC / total AUC, with the trough located by automated peak detection), the counterfactual excess fraction (positive AUC above the log-linear inter-peak decline), rebound height, and freb, an onset-detection-based rebound fraction. Recirculation phenotypes were likewise assigned algorithmically: profiles ranked by the conventional late-peak prominence ratio and partitioned into tertiles (weak / intermediate / strong, n = 5 each). Because this labeling variable is conventional and non-topological, tests of topological descriptors across classes are *concordance* analyses rather than independent validation.

The source study also reported per-profile ex vivo microbiome-derived MPA reactivation rates (fecal β-glucuronidase conversion of MPAG, μM/h). No numeric data or subject keys were published; the anchors were therefore digitized from the source figures, cross-validated against an independent reading, and the profile correspondence reconstructed by joint greedy matching of digitized (AUC, $C_{12}$) pairs against computed values (the adjudicated pairing transposes profiles 10 and 13 relative to digitization order; both pairings and validation statistics in Supplementary Table S4). Anchor analyses are labeled exploratory throughout. Full details of digitization, matching, and reproducibility are given in Supplementary Tables S1 and S4 and Code S1; all 135 descriptor values were independently reproduced by a reimplementation of the pipeline using the gudhi library, and all code and data are publicly available at https://github.com/yxgu2353/TopoPK.

### 4.2 The 15 profiles: universal rebound, heterogeneous shape

The 15 profiles spanned $C_{max}$ 7.6–17.1 mg/L, $AUC_{0-12}$ 27.4–69.9 mg·h/L, and predose $C_0$ 0.77–6.04 mg/L (Table 6). Every profile exhibited an identifiable late local maximum 6.0–11.0 h post-dose (median 10.0 h), with rebound heights of 1.5–7.1 mg/L above the inter-peak trough—consistent with MPA's reported 6–10 h secondary-peak window [38,39]. The EHC AUC fraction ranged 17.3–60.3%: near-universal recirculation varying more than threefold across patients. Fig 6 shows all 15 normalized profiles ordered by Pers_Ratio: the phenotypic axis runs from isolated secondary peaks, through shoulders, to non-isolated late structure.

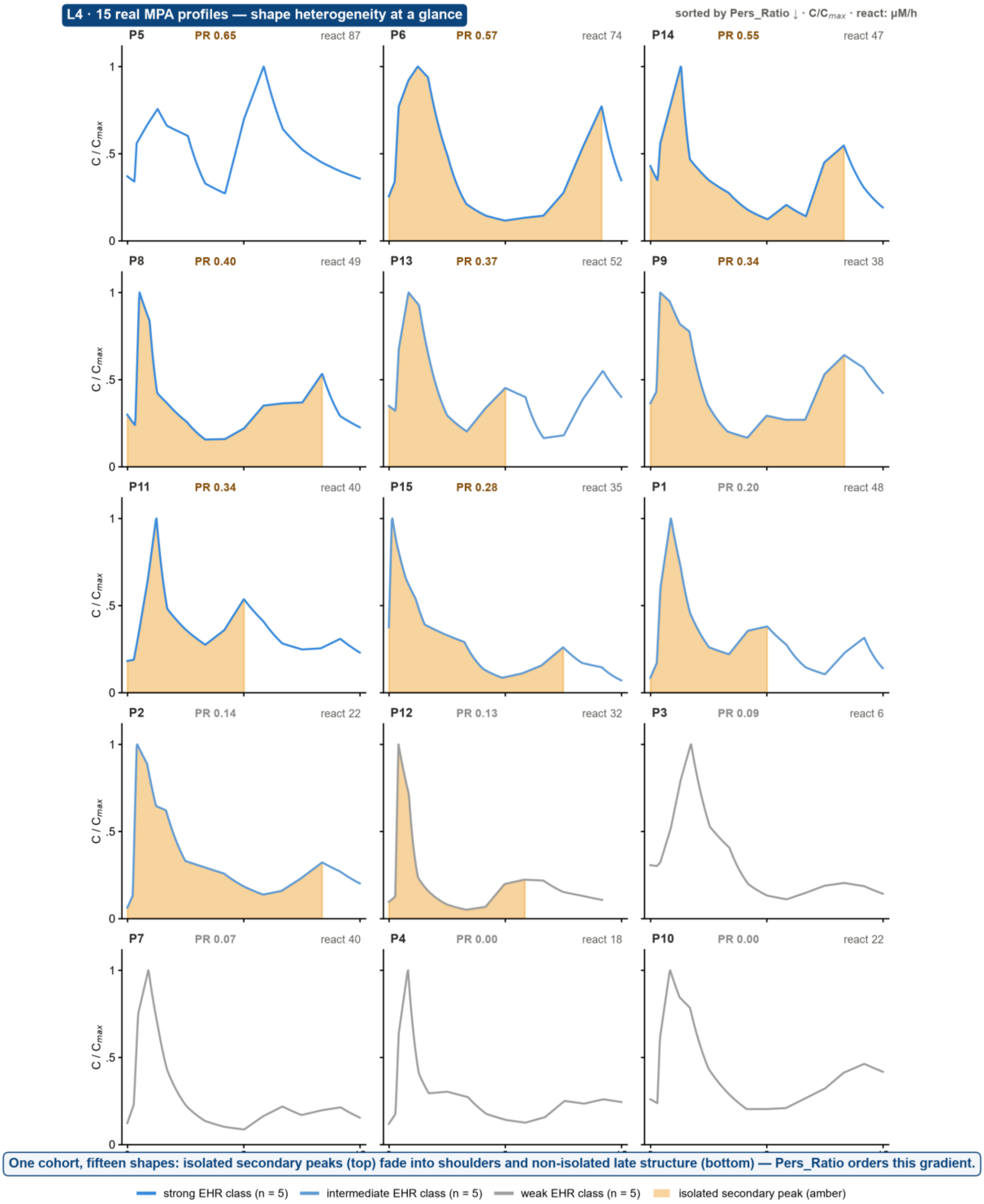

**Fig 6. The 15 real MPA profiles at a glance: one cohort, fifteen shapes.** Each mini-panel: one digitized profile, normalized to C/$C_{max}$ (so shape is the story), colored by its conventional tertile recirculation class (weak gray / intermediate mid-blue / strong blue), with isolated secondary peaks shaded amber. Panels are sorted by Pers_Ratio (descending): isolated secondary peaks (top row) fade into shoulders and non-isolated late structure (bottom row)—Pers_Ratio orders this phenotypic gradient. Header per panel: profile ID, Pers_Ratio, and ex vivo microbiome reactivation rate.

**Table 6.** The 15 MPA profiles—(a) conventional NCA metrics and recirculation operationalizations.

| Profile | Class | $C_{max}$ (mg/L) | $T_{max}$ (h) | $C_0$ (mg/L) | $AUC_{0-12}$ (mg·h/L) | EHC fraction (%) | Rebound (mg/L) | freb (%) | EHR% (manual) | Reactivation (µM/h) |
|---|---|---|---|---|---|---|---|---|---|---|
| 1 | Intermediate | 9.85 | 1.06 | 0.83 | 37.6 | 17.3 | 2.06 | 51.4 | 48.3 | 47.6 |
| 2 | Intermediate | 12.40 | 0.49 | 0.77 | 48.5 | 29.3 | 2.28 | 29.3 | 29.3 | 22.3 |
| 3 | Weak | 15.81 | 2.10 | 4.82 | 60.4 | 22.0 | 1.48 | 22.2 | 21.4 | 6.5 |
| 4 | Weak | 13.92 | 0.99 | 1.64 | 44.9 | 32.9 | 1.72 | 33.2 | 33.7 | 17.7 |
| 5 | Strong | 9.68 | 7.04 | 3.57 | 64.3 | 60.3 | 7.06 | 60.6 | 60.8 | 86.9 |
| 6 | Strong | 10.61 | 1.49 | 2.70 | 53.6 | 41.3 | 6.97 | 41.3 | 42.5 | 73.8 |
| 7 | Weak | 17.08 | 1.09 | 2.08 | 52.6 | 34.7 | 2.25 | 34.8 | 34.8 | 39.9 |
| 8 | Strong | 16.90 | 0.62 | 5.02 | 69.9 | 59.3 | 6.38 | 59.8 | 53.2 | 48.7 |
| 9 | Intermediate | 11.06 | 0.50 | 4.02 | 60.5 | 52.4 | 5.25 | 52.2 | 51.2 | 38.2 |
| 10 | Weak | 7.74 | 1.01 | 2.00 | 37.3 | 45.2 | 2.00 | 45.5 | 41.3 | 22.4 |
| 11 | Strong | 11.67 | 1.50 | 2.10 | 52.2 | 58.6 | 3.03 | 58.6 | 51.9 | 39.7 |
| 12 | Weak | 15.82 | 0.49 | 1.51 | 34.3 | 49.3 | 2.73 | 49.3 | 46.2 | 31.9 |
| 13 | Intermediate | 7.60 | 1.01 | 2.65 | 37.9 | 55.4 | 1.90 | 55.5 | 55.8 | 52.0 |
| 14 | Strong | 14.20 | 1.58 | 6.04 | 60.3 | 42.1 | 5.97 | 42.1 | 42.7 | 47.3 |
| 15 | Intermediate | 8.79 | 0.16 | 3.25 | 27.4 | 29.8 | 1.52 | 29.8 | 30.8 | 34.5 |

**Table 6.** (continued) (b) The nine PTI descriptors.

| Profile | N_H0 | $\beta_1$ | N_PTP | Pers_main | Pers_Ratio | Integral_PI | Entropy_PE | $\Lambda_{NL}$ | Dev |
|---|---|---|---|---|---|---|---|---|---|
| 1 | 3 | 3 | 9 | 0.894 | 0.195 | 1.228 | 1.115 | 0.193 | 0.009 |
| 2 | 2 | 1 | 12 | 0.862 | 0.139 | 0.982 | 0.535 | 0.342 | 0.021 |
| 3 | 2 | 2 | 10 | 0.699 | 0.088 | 0.760 | 0.406 | 0.072 | 0.043 |
| 4 | 1 | 2 | 11 | 0.875 | 0.000 | 0.875 | -0.000 | 0.252 | 0.031 |
| 5 | 2 | 5 | 8 | 0.643 | 0.648 | 1.061 | 0.967 | 0.078 | 0.013 |
| 6 | 2 | 3 | 7 | 0.745 | 0.570 | 1.169 | 0.945 | 0.154 | 0.014 |
| 7 | 2 | 1 | 7 | 0.878 | 0.072 | 0.941 | 0.356 | 0.223 | 0.150 |
| 8 | 2 | 2 | 3 | 0.761 | 0.403 | 1.068 | 0.865 | 0.385 | 0.030 |
| 9 | 2 | 3 | 10 | 0.636 | 0.342 | 0.853 | 0.819 | 0.288 | 0.035 |
| 10 | 1 | 2 | 8 | 0.763 | 0.000 | 0.763 | -0.000 | 0.192 | 0.032 |
| 11 | 3 | 2 | 4 | 0.770 | 0.339 | 1.092 | 1.080 | 0.137 | 0.023 |
| 12 | 2 | 1 | 6 | 0.904 | 0.128 | 1.020 | 0.511 | 0.410 | 0.058 |
| 13 | 3 | 5 | 11 | 0.679 | 0.367 | 1.076 | 1.302 | 0.155 | 0.020 |
| 14 | 3 | 4 | 13 | 0.652 | 0.546 | 1.072 | 1.208 | 0.152 | 0.005 |
| 15 | 2 | 1 | 14 | 0.625 | 0.279 | 0.799 | 0.757 | 0.012 | 0.031 |

The PTI vector's internal structure in these data is low-dimensional: secondary-structure descriptors were strongly intercorrelated (Pers_Ratio vs. Entropy_PE: $\rho = 0.81$, $p = 0.0003$; Entropy_PE vs. N_sig: $\rho = 0.70$, $p = 0.004$), and principal component analysis of the nine descriptors yielded two dominant components (46.1% + 23.3%; three components cumulative 78.8%)—an effective dimensionality of two to three, not nine, to be accounted for in future applications.

## 4.3 Persistence captures peak isolation, not rebound height

Two profiles with visible late rebounds (profiles 4 and 10; rebound 1.7–2.0 mg/L) received Pers_Ratio = Entropy_PE = 0: their secondary features did not survive as persistent H0 bars. Profile 15, with a *smaller* absolute rebound (1.5 mg/L), retained Pers_Ratio = 0.279 (Fig 7, boundary rows). The persistence of a secondary H0 feature depends on the depth of the adjacent minima, not on rebound height alone: profile 10's late concentrations rise gradually from a high shoulder (shallow persistence), whereas profile 15's secondary bump rises from a deep trough (persistent bar). Persistence therefore encodes *peak isolation*—structure invisible to scalar rebound metrics and manual-onset EHR fractions. Because adjacent-minima depth is exactly the quantity most vulnerable to digitization error, this observation awaits confirmation on primary data.

## L4 · Clinical proof of concept — real MPA data

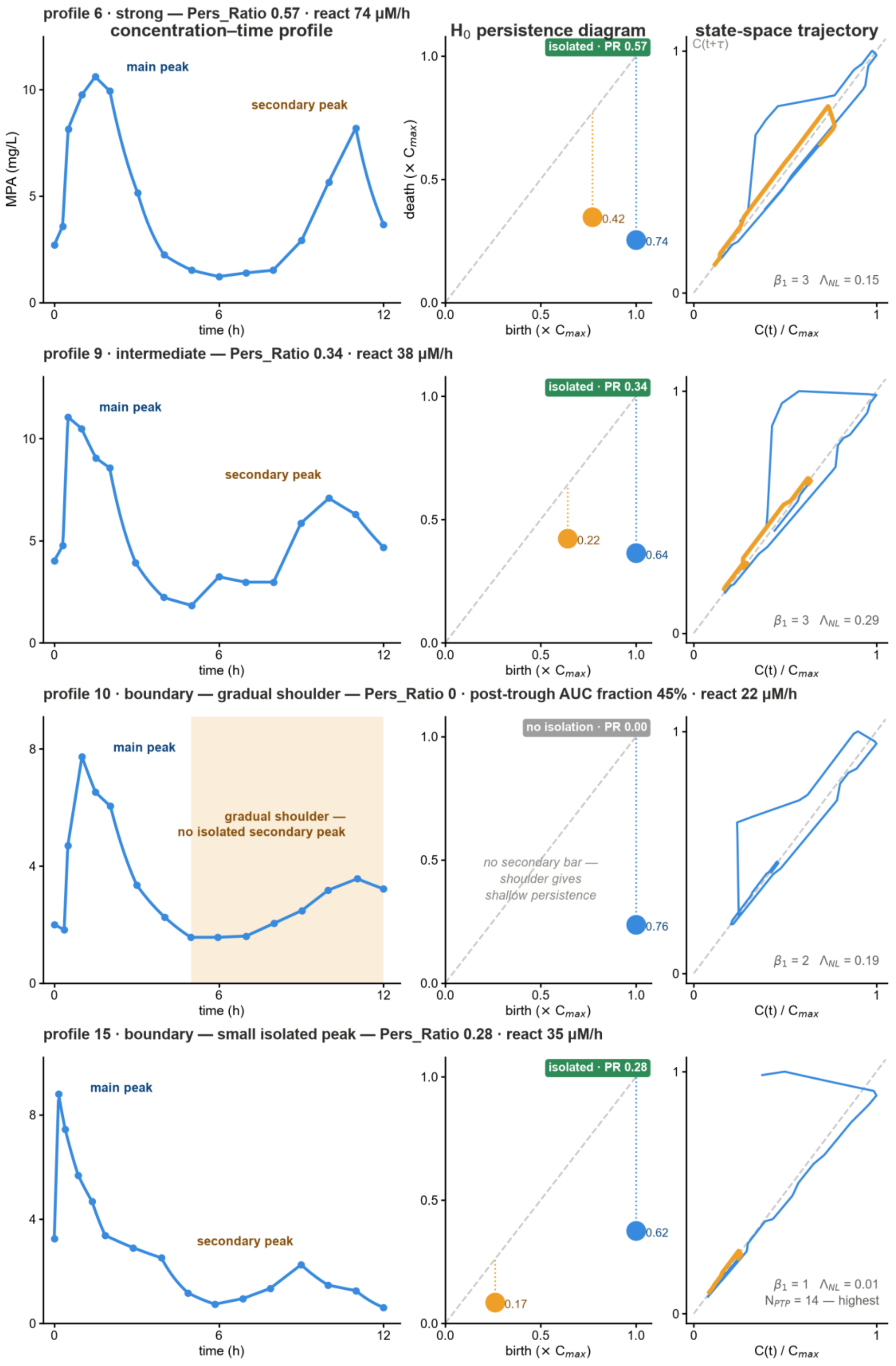

**P10 vs P15 — comparable rebound heights, opposite Pers_Ratio: persistence encodes peak ISOLATION, not rebound height.**

**Fig 7. Topological signatures: persistence encodes peak isolation, not rebound height.** Four representative profiles (rows: strong, intermediate, and two boundary morphologies) × three views (columns: concentration–time profile; H0 persistence diagram; state-space trajectory). The boundary rows are the argument: profile 10's late rise is a gradual *shoulder*—no persistent secondary bar, Pers_Ratio = 0; profile 15's *smaller* rebound rises from a deep trough—a persistent bar, Pers_Ratio = 0.28. Comparable late rises, opposite persistence: the depth of adjacent minima, not height, determines what survives the filtration.

## 4.4 Concordance with conventional metrics

**Concordance where conventional metrics are defined (Fig 8).** Pers_Ratio was associated with the algorithmic EHC AUC fraction (Spearman ρ = 0.53, p = 0.043, nominal; 95% CI 0.06–0.82) and, more strongly, with rebound height (ρ = 0.73, p = 0.002; 95% CI 0.26–0.96)—the only association among nine descriptors across four outcome operationalizations to survive both per-family Bonferroni correction and Benjamini–Hochberg FDR control (q = 0.017). Descriptively, Pers_Ratio and Entropy_PE increased across the algorithmic tertile phenotypes, whereas the AUC-based EHR fractions did not show a corresponding monotone pattern. Because the same conventional late-peak prominence variable defined the groups, panel a is a same-source concordance visualization; Kruskal–Wallis inference is not used as independent validation. Post-trough AUC reflects trough timing as much as rebound magnitude. Topology thus concords with conventional recirculation metrics where they are well defined, and keeps ordering the phenotypes where they do not.

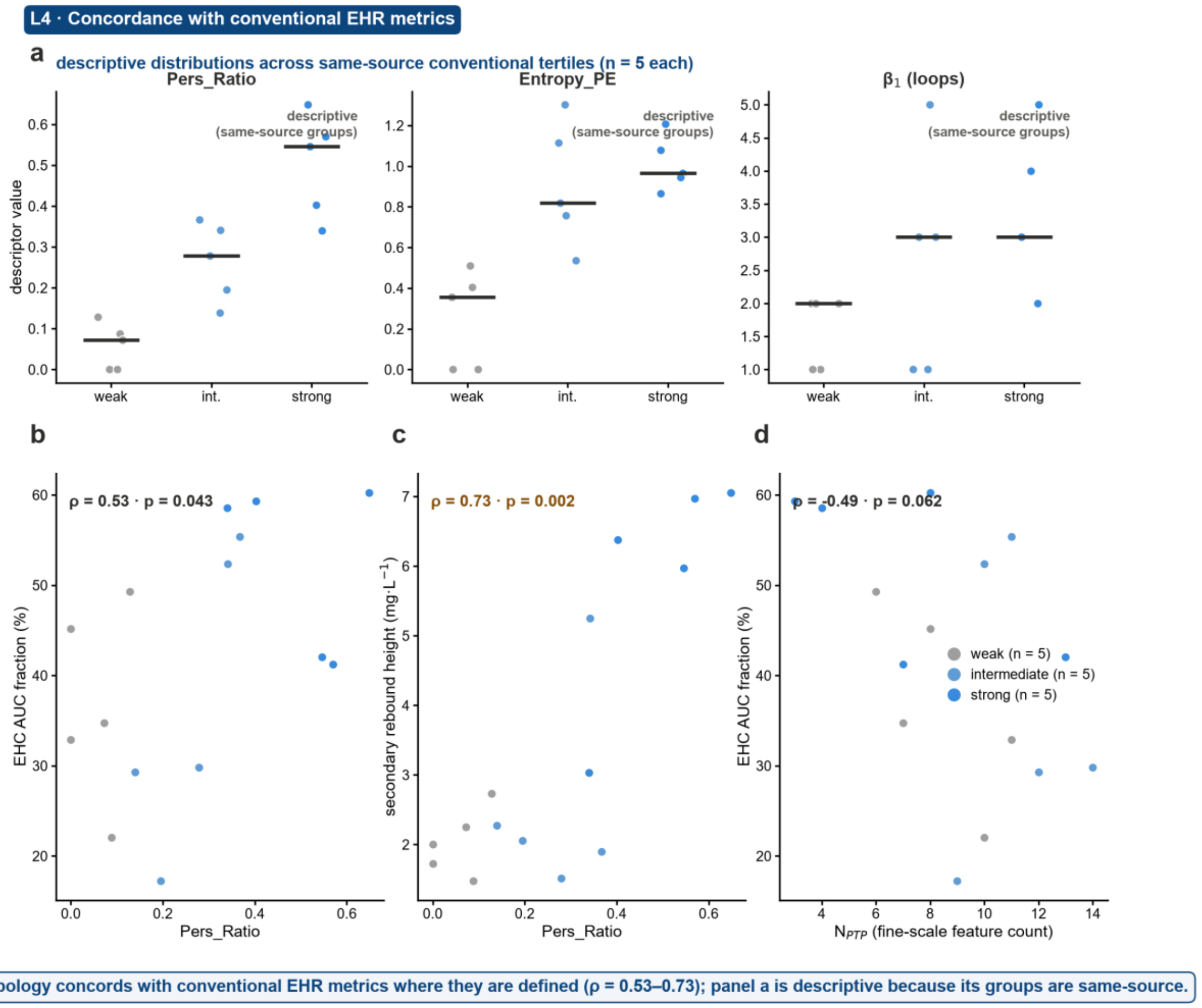

**Fig 8. Concordance with conventional recirculation metrics: pairwise associations plus descriptive ordering of same-source phenotypes.** (a) Descriptive descriptor distributions across algorithmically assigned tertiles (n = 5 each). These groups are defined by conventional late-peak prominence, so this is a same-source concordance visualization rather than independent validation; no Kruskal–Wallis p values are displayed. (b–d) Pairwise concordance: Pers_Ratio vs EHC AUC fraction (ρ = 0.53, p = 0.043); Pers_Ratio vs rebound height (ρ = 0.73, p = 0.002—the headline concordance); N_PTP vs EHC AUC fraction (ρ = −0.49, p = 0.062—negative transfer of the noise-exposed fine-scale count, retained as an honest result).

A complementary honest negative: N_PTP, the fine-scale curvature count, was negatively associated with every outcome operationalization (EHC AUC fraction ρ = −0.49, p = 0.062, nominal) and was highest in the flattest profiles (profile 15: N_PTP = 14). Short bars from digitization noise and low-amplitude wiggle evidently dominate this count under sparse, irregular sampling; fine-scale counting descriptors are the pipeline's most noise-exposed component (consistent with the noise-gradient behavior of Section 3.4) and require persistence floors or stability selection before clinical use.

### 4.5 The microbiome anchor

**The microbiome anchor (Fig 9; Table 7).** Against the ex vivo reactivation rate, Pers_Ratio showed the strongest descriptor association (ρ = 0.815, p = 0.0002; 95% CI 0.48–0.96; leave-one-out ρ = 0.77–0.89), with Entropy_PE (0.744), Integral_PI (0.736), N_sig (0.644), and Dev (−0.600) also surviving within-family correction (q = 0.002–0.032; five of nine descriptors). Reactivation was unrelated to exposure ($AUC_{0-12}$ ρ = 0.35; $C_{12}$ ρ = 0.38). Three observations pin down where the information lives:

1. **The association persists after adjustment for AUC** (rank-partial ρ = 0.79, p = 0.0004): the association remained after AUC adjustment, although this does not exclude all exposure-related confounding.

2. **Conventional associations are attenuated after adjustment for Pers_Ratio** (EHC fraction 0.51 → 0.17, p = 0.56; manual EHR% 0.75 → 0.52)—whereas controlling for the manual EHR% leaves Pers_Ratio's partial association intact (0.656, p = 0.011). The full partial-correlation matrix is given in Supplementary Table S7. This exploratory asymmetric pattern is consistent with Pers_Ratio capturing shared and additional recirculation-related information; with 15 profiles and reconstructed pairing, it does not establish conditional independence or causal information flow.

3. **Sensitivity analyses did not materially weaken the association**: preserved under the worst-case effective-n = 10 (Fisher z p = 0.0025), stable across all 25 embedding-parameter combinations (ρ = 0.815–0.822), and intact after excluding both adjudicated profiles (ρ = 0.780).

Together with the source study's reactivation–EHR% correlation, these results are consistent with a three-way association among gut β-glucuronidase function, recirculation magnitude, and secondary-peak topology, rather than establishing a mechanistic chain. In this dataset, curve shape carried a microbiome-related signal not captured by the tested exposure metrics (ρ = 0.815 vs. ρ ≤ 0.38).

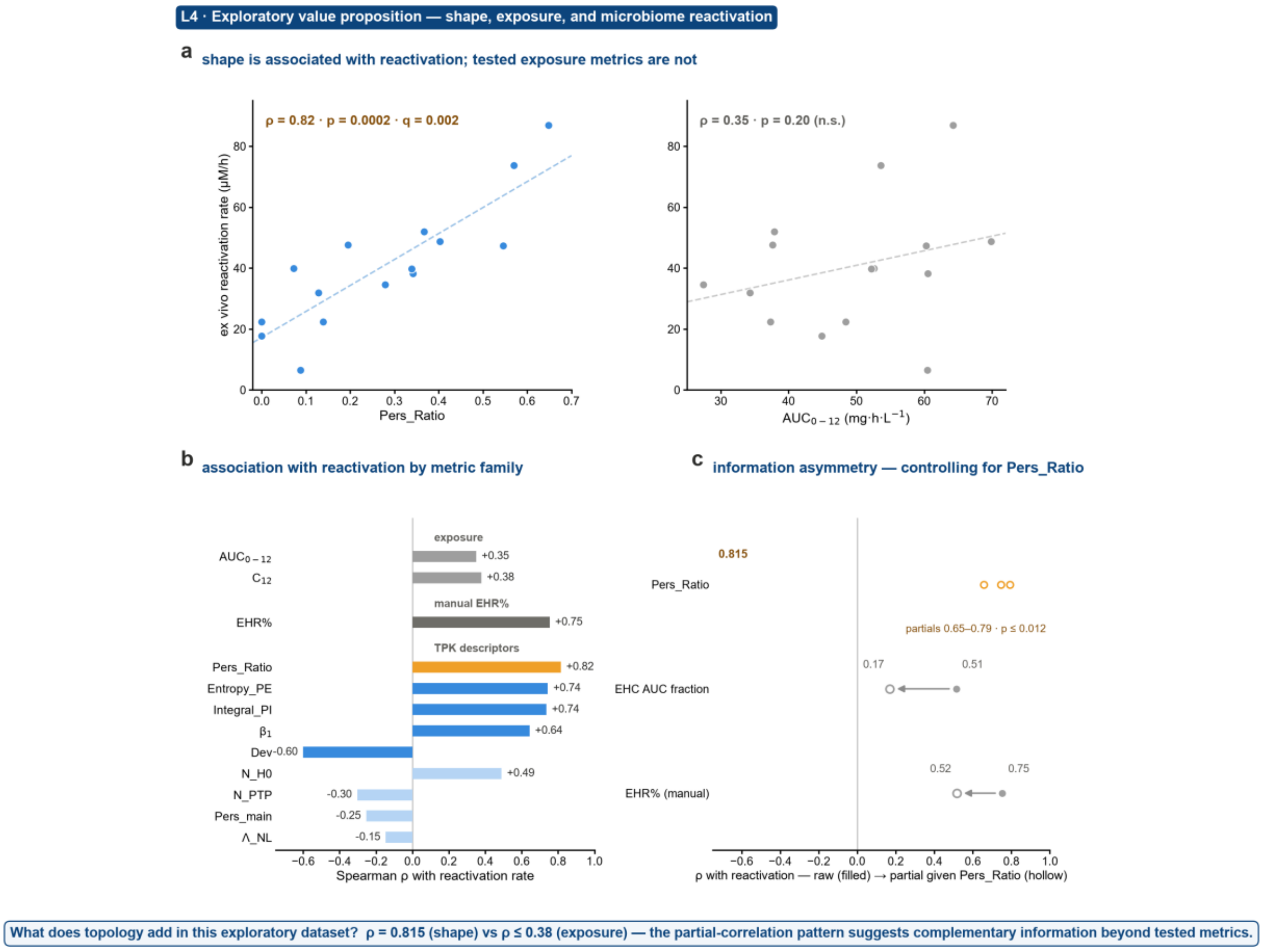

**Fig 9. Exploratory value proposition: shape is associated with gut-microbiome reactivation; the tested exposure metrics are not.** (a) Dual scatter against ex vivo reactivation rate: Pers_Ratio (left, ρ = 0.815, p = 0.0002, q = 0.002) vs $AUC_{0-12}$ (right, ρ = 0.35, p = 0.20, n.s.). (b) Association with reactivation by metric family: exposure metrics (AUC, $C_{12}$; gray, ρ ≤ 0.38), the manually keyed EHR% (ρ = 0.75), and the nine TPK descriptors (five survive within-family correction; Pers_Ratio highest). (c) Information asymmetry: adjustment for Pers_Ratio attenuates the conventional metrics' associations (EHC fraction 0.51 → 0.17; EHR% 0.75 → 0.52), whereas Pers_Ratio's association survives controlling for any of them (partials 0.65–0.79, p ≤ 0.012). Bottom band: the answer, on the canvas—ρ = 0.815 (shape) vs ρ ≤ 0.38 (exposure), and the partial-correlation pattern suggests complementary information beyond the tested conventional metrics.

**Table 7.** MPA–microbiome associations: Spearman ρ, p, and BH-FDR q for the nine descriptors against the ex vivo reactivation rate, with the exposure-metric comparison (all descriptor–anchor pairs in Supplementary Table S6).

| **Metric (family)** | **Spearman $\rho$** | $p$ | **BH $q$ (family)** |
|---|---|---|---|
| Pers_Ratio (TPK) | 0.815 | 0.0002 | 0.002 |
| Entropy_PE (TPK) | 0.744 | 0.0015 | 0.005 |
| Integral_PI (TPK) | 0.736 | 0.0018 | 0.005 |
| N_sig (TPK) | 0.644 | 0.0096 | 0.022 |
| Dev (TPK) | -0.600 | 0.0181 | 0.032 |
| N_H0 (TPK) | 0.490 | 0.0639 | 0.096 |
| N_PTP (TPK) | -0.303 | 0.2725 | 0.350 |
| Pers_main (TPK) | -0.254 | 0.3618 | 0.407 |
| Lambda_NL (TPK) | -0.146 | 0.6025 | 0.603 |
| AUC $_{0-12}$ (exposure) | 0.350 | 0.201 | — |
| C $_{12}$ (exposure) | 0.379 | 0.164 | — |
| EHC fraction (conventional) | 0.514 | 0.050 | — |
| freb (conventional) | 0.675 | 0.006 | — |
| EHR% (manual) | 0.754 | 0.001 | — |

## 4.6 What this application demonstrates

Three points generalize beyond MPA. First, the sampling regime of a standard clinical PK study (5–16 irregular points) falls inside the operating envelope established in Section 3.4—topological shape statistics are computable on routine clinical data, although their interpretation requires external anchoring and further validation. Second, the analysis is fully automated and reproducible: the same fixed pipeline that processed the simulations processed the clinical curves, with no profile-specific operator input in descriptor computation; anchor digitization and pairing required documented manual adjudication. This contrasts with manual EHR quantification, which requires operator-chosen onsets. Third, the entire analysis is a reuse of openly published data: the 15 profiles were digitized from a published figure, and the topological layer revealed an association the original analysis had not extracted. Published pharmacokinetic figures are an under-mined resource; a shape-sensitive, model-free reader adds a layer of questions that can be asked of them without new experiments.

**Why not compartmental refitting—and what happened when we tried.** Each of the 15 profiles was individually refitted with two standard compartmental structures: an enterohepatic model (first-order absorption, two distribution compartments, and a gallbladder compartment discharging once per dosing cycle; eight parameters) and a double-site absorption model (the classical alternative mechanism for secondary peaks; eight parameters). Both were fitted under steady-state twice-daily superposition, with four random starting points per structure. The outcome (Table 8; per-profile fit parameters in Supplementary Table S5, full parameter vectors in Tables S5a/S5b, the parameter-family association scan in Table S5c, and fitted curves overlaid on the observed data in Supplementary Figure S1) illustrates important obstacles that a fixed shape-descriptor pipeline can avoid. The two mechanisms were empirically indistinguishable on these data: the EHC structure fit better on 8 of 15 profiles, the delayed-absorption structure on 7, and the two structures' fitted shape parameters—recirculated fraction and delayed-site absorption fraction—correlated at $\rho = 0.98$; a secondary peak admits at least two equally adequate mechanistic explanations, so a fitted "EHC parameter" is not identifiable as EHC. The fits were weak (median $R^2$ 0.43 and 0.50 across the two structures; the worst profile, $R^2 = 0.10$, accepted a fitted recirculation of twice the dose per cycle), and near-optimal restarts disagreed about the recirculated fraction by up to 0.43 of a dose. The failure was also directional: both structures systematically under-predicted the primary peak (fitted/observed first-peak height: median 0.55, range 0.23–1.01; 14 of 15 profiles below 0.85, the sole exception being the best-fitting profile, $R^2 = 0.93$), and did so nearly identically profile-by-profile ($\rho = 0.98$ between the two structures' peak ratios)—a shared limitation of the two tested model families, not a mechanism-specific one. Two controls locate the failure in the model family rather than the fitting: refitting on the pre-secondary window alone ($t \leq 4$ h) moved the peak ratios by at most 0.07, and refits with twice the random starts and a sixfold iteration budget left $R^2$ and the peak ratios unchanged ($\Delta R^2 \leq 0.002$). A sharp primary peak rising over—and falling back to—a substantial steady-state baseline is precisely the shape a fixed exponential family under twice-daily superposition does not express. Most decisively, the fitted fractions showed no significant association with the microbiome anchor (EHC fraction $\rho = 0.41$, $p = 0.12$; delayed-site fraction $\rho = 0.41$, $p = 0.13$; neither survives multiplicity correction across the parameter family; full scan in Supplementary Table S5c)—ranking below even the crude nonparametric AUC ratio they are meant to formalize ($\rho = 0.51$). After adjustment for Pers_Ratio, the fitted-fraction association was small (partial $\rho = 0.10$, $p = 0.73$), whereas Pers_Ratio remained associated after adjustment for fitted fraction (partial $\rho = 0.77$, $p = 0.001$); this exploratory asymmetry does not establish information flow. The dual-structure refit cost ~24 s per profile; the full topological pipeline costs 0.2 s.

The two conventional poles are thus both accounted for in-data: the NCA pole directly, the compartmental pole by this negative result—complemented by the simulation layers, where the same pipeline reads, under ground truth, exactly the shapes compartmental structures generate, including the gallbladder-cycling EHC model (Sections 3.1–3.3). More flexible structures (multiple emptying events, transit chains, population priors) may improve fit, but would require additional data and careful assessment of identifiability and comparability. Where dense sampling and population data exist, population compartmental fitting with an explicit EHC compartment remains the reference standard; the present regime—sparse, single-occasion, digitized curves—makes individual mechanistic fitting particularly difficult. The topological layer supplies shape descriptors at fixed cost instead: no initialization, no model selection, no fitting; approximately 0.2 s per profile (interpolation, embedding, persistent homology, and all nine descriptors), with the complete 15-profile analysis—including the 25-combination parameter-sensitivity grid—finishing in about a minute.

**Table 8.** Four methods, one question: which quantification of the MPA concentration–time curve tracks the gut-microbiome reactivation rate (n = 15)?

| Method | Quantity | Spearman ρ vs reactivation | Status | Cost per profile |
|---|---|---|---|---|
| NCA (exposure) | $AUC_{0-12}$ | 0.35 (p = 0.20, n.s.) | objective, automated; shape-blind | seconds |
| NCA (exposure) | $C_{12}$ | 0.38 (p = 0.16, n.s.) | objective, automated; shape-blind | seconds |
| Manual EHR% (source study) | operator-keyed secondary-peak AUC ratio | 0.754 (p = 0.001) | carries the signal; operator-dependent, profile-specific onset choice | manual, per profile |
| Compartmental fit (EHC structure) | fitted recirculated fraction | 0.414 (p = 0.12, n.s.; partial given Pers_Ratio 0.10) | mechanism-labeled but structurally ambiguous under this sampling regime | ~24 s (two structures × four starts) |
| Compartmental fit (DSA structure) | fitted delayed-site fraction | 0.411 (p = 0.13, n.s.) | same | (included above) |
| TPK (this work) | Pers_Ratio | 0.815 (p = 0.0002, q = 0.002) | fully automated, fixed pipeline, mechanism-agnostic | 0.2 s |

Compartmental rows: individual steady-state refitting with eight-parameter enterohepatic and double-site-absorption structures (four random starts each; per-profile parameters in Supplementary Table S5). Exposure and manual rows as analyzed in Sections 4.1–4.5. The exposure metrics are shape-blind; the operator-keyed ratio carries the signal but is subjective and profile-specific; the compartmental parameters are objective in form but structurally ambiguous at this sampling density; the topological descriptor is the only quantification that is simultaneously objective, mechanism-agnostic, and strongly associated.

In the terms of this section's opening problem: a model-free, fully automated reader recovered from these 15 curves a recirculation signal that was stronger than those obtained from exposure metrics or compartmental refits and comparable to the manual EHR% ratio, while avoiding its profile-specific onset selection. The analysis required seconds of computation and no profile-specific operator decisions.

## 5. Discussion

The investigations reported here constitute a progressive evaluation of a central hypothesis: that curve shape, read through persistent homology, can carry information not captured by the tested scalar summaries. The evidence stands on four pillars. In linear systems the method reports the absence of cyclic structure where none exists (N_sig = 0 across the four monophasic linear benchmarks). Under nonlinearity it resolves curvature-without-cycles, and separates linear shape rigidity from dose-dependent deformation across 24 scenarios. Beneath an exogenous drug it detects a weak endogenous rhythm whose shape distortion survives superposition exactly (D_dev = 1.410 = 1.410). And on real clinical data, the secondary-peak topology of MPA profiles was associated with gut-microbiome reactivation at ρ = 0.815 where the tested exposure metrics show weak, nonsignificant associations (ρ ≤ 0.38)—with an exploratory asymmetric partial-correlation pattern consistent with additional information beyond the tested conventional metrics.

**Theoretical significance.** TPK's defining capability is recovering dynamical information from a one-dimensional observable. The concentration–time curve is a projection of the full ADME manifold; delay embedding reconstructs a point cloud capturing its prominent cyclic and curved structure, and persistent homology reads its invariants. What is recovered is not the full geometry—that would require observing all state variables—but its topological essence: the loops, components, and persistent hierarchies that survive projection. This is a mathematically principled distillation that operates orthogonal to conventional statistical compression: it retains complementary multiscale structure rather than replacing the curve with a single scalar; persistence-diagram perturbation stability is formalized in the bottleneck metric [14,15], while descriptor-level robustness remains an empirical question. The simulation layers also demonstrated the *diagnostic specificity* of this reduction: curvature and cycles are cleanly separated (nonlinear elimination bends, rhythms loop), and DSA's coupled parallel peaks are separated from EHC's sequential physiological cycling by decoupling and interval alone.

**What the clinical application adds.** The MPA analysis is, to our knowledge, the first application of persistent-homology shape descriptors to real clinical concentration–time data. Three findings stand out. First, the secondary-structure descriptors anchored the microbiome-derived reactivation rate (ρ = 0.815, q = 0.002), with the observed association residing in shape rather than the tested exposure summaries: adjustment for AUC left the association similar, whereas adjustment for Pers_Ratio attenuated the conventional metrics' associations. Second, persistence encoded *peak isolation*—a multi-scale property invisible to scalar rebound metrics: two profiles with equal rebound heights received Pers_Ratio of 0 and 0.28 depending on whether the secondary feature rose from a shoulder or from a deep trough. Third, the entire quantification was algorithmic and reproducible—the same fixed pipeline that processed the simulations processed the clinical curves—whereas the reference EHR quantification depends on an operator-chosen onset sample per patient (for one profile, the manual and algorithmic onset choices differ by 31 AUC-percentage points). No approach is threshold-free; the distinction is that TPK's choices are few, explicit, uniformly applied, and—crucially—multi-scale, distinguishing shallow shoulders from isolated peaks where any single-threshold metric conflates them.

**AI-ready pharmacokinetics.** The PTI vector transforms the concentration–time curve from a model-dependent, irregularly sampled time series into a fixed-length vector in a geometrically meaningful shape space. In this space, pharmacokinetically similar dynamics cluster regardless of dose amplitude or route. This is the prerequisite for large-scale machine learning on pharmacokinetic data: a standardized, dose-normalized, model-agnostic feature set. The same property makes published pharmacokinetic figures—an under-mined, openly licensed resource—systematically re-interrogable: the present clinical layer is itself a demonstration, extracting an association the original study's coordinate system could not express.

**From shape to regimen and response.** The developmental payoff of a cheap, reproducible shape reading runs through exposure–response analysis. Shape is the temporal structure of exposure: for narrow-therapeutic-index drugs, and for drugs whose pharmacodynamics is itself path-dependent—hysteresis, tolerance, rebound—the timing of concentration excursions, not only their integrated size, drives both effect and toxicity. The MPA case contains this chain in miniature: the same gut-microbial deconjugation that restores systemic immunosuppressive exposure [10] also returns active drug to the gut lumen, long implicated in MPA's dose-limiting gastrointestinal toxicity [38,39]—one edge of the graph, read from both sides. In current practice that temporal structure is either averaged away (AUC) or gated behind compartmental fitting (Section 4.6); TPK makes it a directly computable covariate. Three concrete uses follow. (i) *Exposure–response analysis:* PTI descriptors enter as candidate covariates alongside AUC, $C_{max}$, and trough—cheap screening tools for the exposure metric that best predicts response or toxicity in a given program. (ii) *Regimen design:* an individual recirculation phenotype bears directly on decisions that are otherwise population-averaged—dosing frequency, food state, co-administration with bile-acid sequestrants (the classical recirculation-interrupting interaction [36]), and the intensity of therapeutic drug monitoring. (iii) *Early-development triage:* a fully automated, 0.2-s-per-curve sentinel that flags nonlinear elimination, rhythmic interference, or anomalous secondary structure in first-in-human sparse data—selecting which compounds and which subjects warrant model-based depth. TPK does not predict effect by itself; it widens the vocabulary in which exposure is described, so that efficacy and toxicity—the ultimate endpoints—can be regressed against the temporal structure that generated them.

**Limitations.** Several boundaries apply. (i) The clinical data were digitized from published figures; digitization was cross-validated (independent extraction, pixel-level re-verification of the anchor pairing), and the source study's published statistics were reproduced, but confirmation on subject-keyed primary data is required. (ii) The sample is small—15 profiles from 10 patients, pairing unavailable—so p values are anticonservative; the headline association nevertheless survived a worst-case effective-n = 10 analysis. (iii) The embedding thresholds (prominence 5% of $C_{max}$, H1 persistence 0.015) are empirical; sensitivity grids covered all claimed associations (MPA grid in Supplementary Table S2), and the operating-characteristics study (Section 3.4) maps the envelope, but alternative filtrations and metrics were not explored. (iv) Delay embedding under 5–16 irregular samples satisfies none of the hypotheses of Takens' theorem; embedding-based features are therefore interpreted as shape statistics of the observed curve, not as reconstructions of an underlying attractor. (v) One descriptor class failed to transfer from simulation to clinic: fine-scale curvature counts (N_PTP) were dominated by digitization noise under sparse sampling—the pipeline's noise-exposed extreme, consistent with the noise-gradient behavior of Section 3.4, and requiring persistence floors or stability selection before use. (vi) This is a single drug, dose, and population; MPA's pronounced EHR is a favorable test case, and drugs with subtler shape features may demand denser sampling. (vii) The Vietoris–Rips filtration scales quadratically with series length, constraining very long series without subsampling.

**Toward topological pharmacology.** The logic of TPK extends with full symmetry to pharmacodynamics. A pharmacodynamic time series—blood pressure, T-cell activation, biomarker response—is likewise a one-dimensional projection of a high-dimensional response manifold, and drug effect is frequently path-dependent: tolerance loops, rebound dynamics, chronopharmacodynamic modulation are all, in the present vocabulary, shape phenomena. The development of Topological Pharmacodynamics (TPD) is the immediate next step in our program; when PK and PD shape features are placed in a common space, the result is a unified topological pharmacology—a single geometric language describing a drug's fate and its effect.

Pharmacokinetics has spent half a century asking *how much*—how much drug is in the blood, how large the exposure, how fast the clearance. Topological Pharmacokinetics asks *what shape*. In the curve of concentration over time there is geometry: loops where physiology cycles, bends where kinetics saturate, structures where endogenous rhythms persist. This geometry has always been there, encoded in the data, invisible to tools designed to compress curves into numbers. These results show that parts of it can be read systematically, while external validation remains necessary.